\documentclass[pdflatex,sn-mathphys-num]{sn-jnl}

\usepackage{graphicx}%
\usepackage{multirow}%
\usepackage{amsmath,amssymb,amsfonts}%
\usepackage{amsthm}%
\usepackage{mathrsfs}%
\usepackage[title]{appendix}%
\usepackage{xcolor}%
\usepackage{textcomp}%
\usepackage{manyfoot}%
\usepackage{booktabs}%
\usepackage{algorithm}%
\usepackage{algorithmicx}%
\usepackage{algpseudocode}%
\usepackage{listings}%
\usepackage{subfig}

\renewenvironment{table}[1][]%
{\tableorg[#1]%
	\tablebodyfont%
	\renewcommand\footnotetext[2][]{{\removelastskip\vskip3pt%
			\let\tablebodyfont\tablefootnotefont%
			\hskip0pt\if!##1!\else{\smash{$^{##1}$}}\fi##2\par}}%
}{\endtableorg}

\usepackage{float}
\usepackage{array}

\theoremstyle{thmstyleone}%
\theoremstyle{thmstyletwo}%

\theoremstyle{thmstylethree}%

\begin{document}
	
	\title[Article Title]{Energy-Efficient Real-Time 4-Stage Sleep Classification at 10-Second Resolution}
	
	
	\author[1]{\fnm{Zahra} \sur{Mohammadi}}\email{zahramohammmadi@ut.ac.ir}
	
	\author[1]{\fnm{Parnian} \sur{Fazel}}\email{parnian.fazel@ut.ac.ir}

	\author*[1]{\fnm{Siamak} \sur{Mohammadi}}\email{smohamadi@ut.ac.ir}

	\affil[1]{\orgdiv{School of Electrical and Computer Engineering}, \orgname{University of Tehran}}

	
	\abstract{Sleep stage classification is critical for diagnosing and managing disorders like sleep apnea and insomnia. However, conventional methods like polysomnography are costly and impractical for long-term, home-based monitoring. This study presents an energy-efficient approach for detecting four sleep stages (wake, rapid eye movement (REM), light sleep, deep sleep) using a single-lead electrocardiogram (ECG) signal. We evaluate various machine learning and deep learning models, introducing two windowing strategies: (1) a 5-minute window with 30-second steps for machine learning and (2) a 30-second window with 10-second steps for deep learning, enabling 10-second temporal resolution for real-time predictions. While deep learning models like MobileNet-v1 achieve high accuracy (92\%) and F1-score (91\%), their energy demands make them unsuitable for wearables. To address this, we design SleepLiteCNN, optimized for ECG-based sleep staging, achieving 89\% accuracy and 89\% F1-score while minimizing energy use. Applying 8-bit quantization further reduces energy consumption to 5.48 $\mu$J per inference, with 90\% accuracy and F1-score. Additionally, field-programmable gate array (FPGA) deployment shows significant reductions in resource usage. This approach provides a practical, energy-efficient solution for continuous ECG-based sleep monitoring in resource-constrained wearable devices.} 

	
	
	
	\maketitle
	
	\section{Introduction}\label{Intro}
	Sleep stage classification is crucial in health monitoring, especially for diagnosing and managing sleep disorders such as sleep apnea and insomnia \cite{1}. Sleep stages are categorized into wake, REM, and Non-Rapid Eye Movement (NREM) stages. NREM is further divided into four stages: N1, N2, N3, and N4. N1 and N2 represent light sleep, during which the body begins to relax, while N3 and N4 correspond to deep sleep, which is essential for physical recovery and memory consolidation \cite{2,3}.
	
	Polysomnography is the gold standard for sleep staging, usually performed in specialized clinics. It measures multiple physiological signals such as electroencephalogram (EEG), ECG, blood oxygen saturation (SpO2), and snoring \cite{4}. Although polysomnography is accurate and comprehensive, it is costly, time-consuming, and often uncomfortable for patients, making it unsuitable for long-term or home-based monitoring. As a result, there is a growing demand for cost-effective, user-friendly alternatives based on wearable devices. Although EEG-based approaches have been widely explored due to their direct correlation with neural activity, they require multiple scalp electrodes, making them inconvenient for daily use.
	
	ECG signals provide a more practical alternative, as they can be recorded using portable and unobtrusive devices with fewer electrodes---making them well-suited for continuous sleep monitoring \cite{2,9,10}. While many ECG-based classification studies have achieved reliable performance in 2-stage (wake, sleep) or 3-stage (wake, REM, and NREM) tasks, relatively fewer have demonstrated high performance in 4-stage classification (wake, REM, light sleep, and deep sleep)\cite{2,9,10}. Despite recent advances, no prior ECG-based method has achieved 10-second resolution in 4-stage classification while maintaining energy efficiency suitable for wearable devices.
	
	In wearable systems, shorter temporal resolutions significantly impact a device's ability to provide real-time or near-real-time monitoring. Previous approaches typically offer 30-second or longer temporal resolutions, which may introduce delays in detecting important transitions in sleep stages \cite{2,9,10,6,5}. While machine learning requires longer windows to extract robust handcrafted features, deep learning enables automatic feature extraction, allowing for shorter windows with higher temporal resolution---a critical advantage for real-time wearable systems. By employing a 30-second window with a 10-second step for deep learning, we enable near-real-time (10-second) resolution for sleep stage predictions, resulting in a more dynamic and responsive monitoring system. Higher temporal resolution may also help capture rapid sleep-stage transitions and brief arousals that are difficult to observe using conventional 30-second epochs.
	
	However, the reduced window length demands sophisticated deep learning architectures to maintain classification performance. Meanwhile, machine learning models require larger windows (e.g., a 5-minute window with a 30-second step) to ensure accurate feature extraction, thereby achieving a 30-second resolution \cite{12}.
	The windowing strategies adopted in this work are designed to accommodate the different requirements of machine learning and deep learning models. In particular, the 30-second sliding window with a 10-second step for deep learning enables sleep stage prediction at a finer temporal resolution (10 seconds) while preserving sufficient physiological context for accurate classification. The novelty of our approach lies not in the use of overlapping windows themselves, which are common in signal processing, but in enabling ECG-based 4-stage sleep classification at 10-second resolution together with energy-efficient model deployment for wearable systems.

	Additionally, wearable devices must operate under strict energy and resource constraints. Larger and more complex models, such as deep learning architectures, often result in excessive energy consumption and heat dissipation, which reduce battery life and negatively impact usability. Therefore, energy consumption per inference is a critical factor in ensuring the long-term feasibility of autonomous sleep monitoring systems. In this paper, we assess energy consumption using 45 nm technology to provide realistic energy estimates for practical deployment in wearable applications.
	
	To address these challenges, we introduce a comprehensive pipeline for ECG-based 4-stage sleep classification that effectively balances classification accuracy, temporal resolution, and hardware efficiency. To our knowledge, this is the first ECG-based method to achieve 10-second resolution in 4-stage classification. The key contributions of this work include:
	
	\begin{enumerate}
		\item{\textbf{High-Performance 4-Stage Classification:} Among the deep learning models evaluated, MobileNet-v1 achieves up to 92\% accuracy and a 91\% F1-score in distinguishing wake, REM, light sleep, and deep sleep, with an approximate 5\% improvement in accuracy and a 9\% improvement in F1-score over previous ECG-based methods \cite{2}.}

		\item{\textbf{High-Resolution Sleep Stage Prediction:} By employing a 30-second sliding ECG window with a 10-second step together with lightweight deep learning architectures, our framework enables 4-stage sleep classification at a 10-second temporal resolution. This finer temporal granularity supports near-real-time sleep monitoring in wearable devices while maintaining high classification performance.}
		
		\item{\textbf{Comprehensive Machine Learning and Deep Learning Evaluation:} We systematically evaluate multiple machine learning algorithms and deep learning architectures on the same 4-stage classification task, providing a comprehensive assessment of their effectiveness in ECG-based sleep staging.}
		
		\item{\textbf{Quantization-Aware Training for Energy Efficiency:} To address the energy and resource constraints of wearable devices, we apply 8-bit quantization to deep learning models, significantly reducing inference energy consumption while maintaining accuracy.}
		
		\item{\textbf{SleepLiteCNN Design for Energy-Efficient Inference:} Recognizing that even lightweight deep learning models are computationally expensive, we develop a SleepLiteCNN aimed at reducing resource usage and energy consumption, rather than focusing solely on maximizing performance.}
		
		\item{\textbf{FPGA Resource Utilization of the SleepLiteCNN:} While this study does not advocate FPGA as the sole hardware platform, we evaluate the resource utilization of our SleepLiteCNN on an FPGA to demonstrate its suitability for resource-constrained wearable devices.}
	\end{enumerate}
	
	By integrating short temporal resolutions, energy-efficient model optimization, and comprehensive machine learning and deep learning evaluations, our approach advances the feasibility of energy-efficient, ECG-based sleep staging in wearable applications. The remainder of this paper is organized as follows: Section \ref{Related Works} reviews related work on ECG-based sleep classification, focusing on multi-stage classification and energy efficiency. Section \ref{Methodology} explains our methodology, including data preprocessing, feature extraction, and the windowing strategies used for both machine learning and deep learning models. Section \ref{Results} presents experimental results, including classification performance, energy consumption analysis, and hardware efficiency. Section \ref{Limitations} discusses the limitations of the current study and outlines potential future research directions. Section \ref{Discussion} provides an analysis and interpretation of the results, highlighting their implications and comparison with existing methods. Finally, Section \ref{Conclusion} concludes the paper.

	\section{Related Works}
	\label{Related Works}
	In this section, we review past research on sleep stage classification, focusing on the increasing reliance on ECG signals and the importance of balancing classification performance, temporal resolution, and power efficiency. Despite advances in machine learning and deep learning, many existing approaches still depend on multiple EEG electrodes or give insufficient attention to the energy and resource constraints of wearable devices \cite{19,21,25,45,46,47,48,49}. Wearables demand accurate multi-stage detection, rapid inference, and ultra-low power consumption, making these goals difficult to achieve in practice. To address these challenges, we explore five major themes: (1) ECG-based 4-stage classification, (2) approaches to temporal resolution, (3) traditional machine learning methods, (4) modern deep learning strategies, and (5) energy-efficient architectures and hardware considerations. We conclude by highlighting the key research gaps that persist in this field.
	
	\subsection{ECG-Based Sleep Staging}
	Early sleep staging systems primarily relied on EEG signals, but EEG setups can be inconvenient for daily use. Consequently, researchers have turned to ECG signals, which require fewer sensors and are easier to record. Initially, most ECG studies focused on 2-stage (wake, sleep) or 3-stage (wake, REM, NREM) classifications \cite{24}, \cite{23}, \cite{22}. While much attention has been given to achieving high performance in these classifications, less focus has been placed on improving performance in 4-stage classification (wake, REM, light sleep, and deep sleep). For instance, one study using heart rate variability (HRV) and R-peak features achieved 87.15\% accuracy in 4-stage classification \cite{2}, and a Convolutional Neural Network (CNN) using instantaneous heart rate (IHR) reached 77\% accuracy on large datasets \cite{10}. However, many of these methods still rely on long time windows or do not prioritize power efficiency, making them less suitable for continuous monitoring on wearable devices.
	
	\subsection{Temporal Resolution in Sleep Staging}
	High temporal resolution is essential for real-time sleep staging, particularly for detecting micro-arousals and sudden transitions between sleep stages. Traditional approaches typically rely on 30-second or longer epochs, which, while beneficial for stable feature extraction, may delay system responsiveness and miss brief but clinically significant events. To address this, several studies have explored multi-scale windowing strategies. For instance, a CNN-LSTM model in \cite{16} used 270-second input windows with 30-second steps to improve 5-stage classification, while another study reported 85.3\% accuracy for 4-stage classification using 40.96-second segments across multiple users \cite{23}. Additionally, a recent study in \cite{46} achieved 30-second resolution without overlapping windows using a Bi-LSTM model. SleepContextNet \cite{49} extended this idea by modeling both short-term and long-term temporal context using CNNs and a unidirectional RNN over sequences of 30-second epochs. This approach captures sleep stage transitions across a broader temporal horizon, boosting classification performance without increasing the temporal resolution of input data.
	
	Despite these advances, limited attention has been given in the literature to achieving higher temporal resolution (e.g., 10-second steps), which is critical for near-real-time sleep monitoring in wearable systems. This gap highlights the need for methods that combine fine-grained temporal resolution with efficient computation to support responsive, continuous sleep tracking.
	
	\subsection{Machine Learning Approaches}
	Traditional machine learning approaches for ECG-based sleep staging typically rely on handcrafted features---such as HRV, ECG-derived respiration (EDR), or morphological parameters---and classifiers like support vector machines (SVMs), Bayesian linear discriminants, or tree-based ensembles \cite{5,6,18,22,55}. While these methods have achieved moderate accuracy for 4-stage classification, typically ranging from 70\% to 80\% \cite{5,6}, they often rely on long segments (e.g., 5-minute epochs) to maintain feature stability. Although these methods are generally more interpretable and computationally less intensive than deep learning models, their heavy dependence on manual feature extraction limits their flexibility in dynamic, wearable environments. Furthermore, these approaches rarely incorporate energy-efficient optimizations, which are crucial for continuous overnight monitoring on devices with limited battery life.
	
	\subsection{Deep Learning Approaches}
	Deep learning models are capable of automatically learning complex patterns from raw or minimally processed ECG signals, eliminating the need for extensive handcrafted feature extraction. Architectures such as CNNs, long short-term memory (LSTM) networks, bidirectional LSTMs (biLSTM), and gated recurrent units (GRUs) have shown strong potential for multi-stage sleep classification. For instance, a Deep Convolutional Recurrent (DCR) model achieved 74.2\% accuracy for five-stage classification using single-lead ECG signals \cite{20}, while another study applying an LSTM to HRV data reported 77\% accuracy for 4-stage classification \cite{26}.
	
	\subsection{Energy-Efficient Designs and Hardware Considerations}
	Some research has focused on energy-efficient designs and hardware considerations for automatic sleep staging systems, recognizing the stringent power constraints of wearable devices. For example, \cite{17} demonstrated an ultra-low-power dual-mode processor that leverages an algorithm-hardware co-design approach, achieving sub-10 $\mu W$ power consumption and 0.149 mJ energy consumption in 180 nm technology by employing neural-network based decision trees with aggressive computation pruning and clock gating. Similarly, \cite{44} developed an ultra-low-power system-on-chip that integrates an analog front end with a digital processor optimized for spectral feature extraction, resulting in a total power consumption of 575 $\mu W$ and 10.35 mJ energy consumption in 180 nm technology while maintaining high classification accuracy. Although \cite{15} primarily focuses on enhancing classification performance through a mixed neural network approach, their work underscores the emerging trend of integrating advanced deep learning techniques with hardware-friendly architectures to support real-time, wearable sleep monitoring.
	
	Despite advances in 4-stage ECG-based sleep classification, many methods still rely on long epochs, compromise energy efficiency, or lack the high temporal resolution necessary for real-time monitoring. Even those using shorter epochs often overlook hardware-focused design or fail to track power consumption, highlighting the need for a solution that combines high accuracy, fine-grained resolution, and low power usage, especially in wearables. To meet this need, we propose an energy-efficient pipeline for 4-stage ECG classification that incorporates 10-second resolution, 8-bit quantization, and a SleepLiteCNN optimized for resource-constrained, wearable deployment.

	\section{Methodology}
	\label{Methodology}
\begin{figure}[H]
		\centering
		\includegraphics[width=1\linewidth]{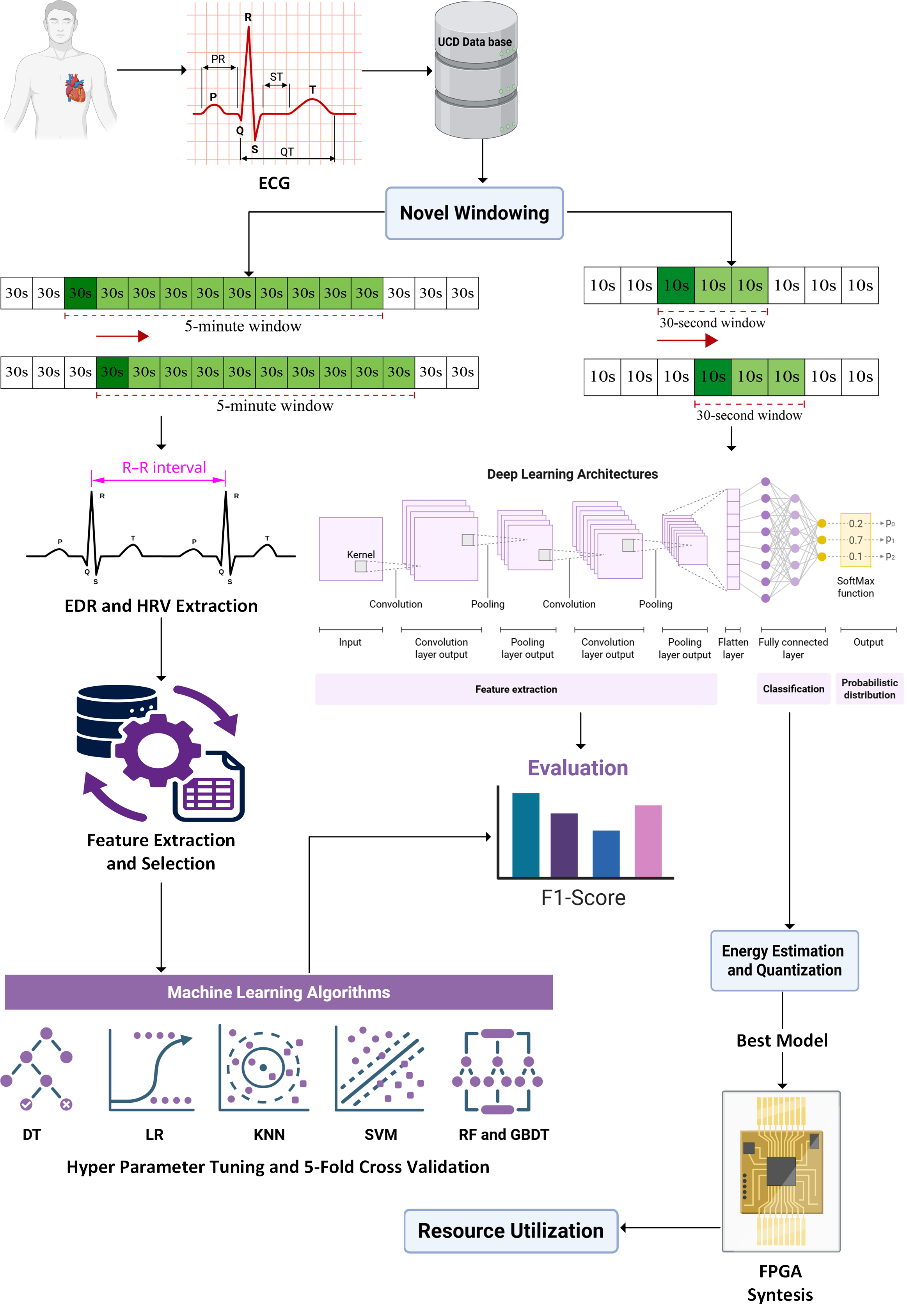}
		\caption{ECG-Based Sleep Staging: Balancing Accuracy, Energy Efficiency, and Temporal Resolution.}
		\label{fig_1}
	\end{figure}

	As shown in Figure~\ref{fig_1}, our methodology prioritizes short temporal resolution and energy efficiency in wearable ECG-based sleep staging. We introduce two windowing strategies: a 5-minute window (with a 30-second step) for machine learning, which captures frequency-domain and nonlinear-domain features, and a 30-second window (with a 10-second step) for deep learning, enabling 10-second resolution and near real-time detection of significant sleep transitions. In the machine learning pipeline, we extract and select handcrafted features prior to training multiple classical algorithms. Although existing lightweight CNNs like MobileNet and SqueezeNet reduce computational load, they still contain numerous parameters and are not specifically optimized for ECG signals. To address these shortcomings, we developed a SleepLiteCNN tailored for ECG-based sleep classification, resulting in a compact design with only 47K parameters. This specialized architecture targets resource-constrained settings, facilitating continuous monitoring on battery-powered devices.
	
	To further minimize energy consumption, we apply 8-bit quantization---a format widely supported on modern hardware---that strikes a practical balance between model size, computational overhead, and accuracy. We then estimate each model's energy usage at the 45 nm technology node and synthesize the most efficient design on an FPGA, confirming its feasibility for continuous sleep monitoring in resource-constrained wearable devices. The subsections below describe each stage of this integrated process, from data preprocessing and windowing to feature extraction, model design, and evaluation.
	
	\subsection{Dataset}
	In this study, we used the UCDDB dataset \cite{7}, which contains 25 full overnight polysomnography recordings (6--8 hours each) from adult subjects suspected of sleep disorders. Each 30-second segment was annotated by sleep experts based on standard sleep staging criteria. For our analysis, we focused on single-lead ECG signals, specifically from the modified lead V2, sampled at 128~Hz. We chose ECG over EEG because it requires fewer electrodes and is therefore easier to implement in wearable devices. 
	
	Our analysis of the UCDDB dataset confirmed no signal loss in the ECG recordings, allowing us to utilize all available signals without requiring gap-filling. Additionally, we opted not to denoise or normalize the ECG signals, enabling our models to learn noise-resilient features and improving their generalization to real-world wearable applications.
	
	After data windowing, we split the dataset into 80\% for training and 20\% for testing. From the training set, we further allocated 10\% as a validation set for use in deep learning model development. We used the test set exclusively for final performance evaluation to avoid data leakage. For the machine learning models, we performed 5-fold cross-validation during Bayesian hyperparameter tuning within the training subset. 
	
	In this study, sleep stages N1 and N2 were merged into a single \textit{Light Sleep} class, and stages N3 and N4 were merged into a \textit{Deep Sleep} class. This consolidation is commonly adopted in automatic sleep staging studies to reduce class imbalance and mitigate the high inter-scorer variability associated with N1, while preserving the main physiological structure of sleep architecture. Moreover, several wearable-oriented sleep monitoring systems employ similar stage grouping to improve classification robustness in practical settings \cite{2,4,10}.

	\subsection{Data Windowing}
\begin{figure*}[!ht]
		\centering
		\subfloat[]{\includegraphics[width=0.6\linewidth]{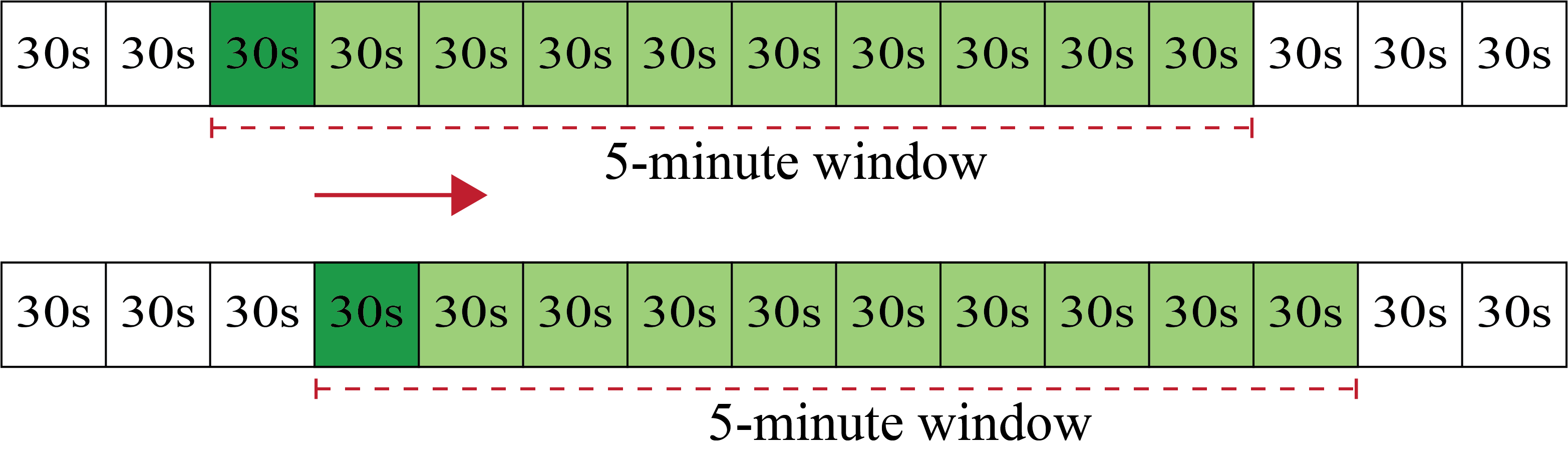}%
			\label{fig_first_case}}
		\hfil
		\subfloat[]{\includegraphics[width=0.32\linewidth]{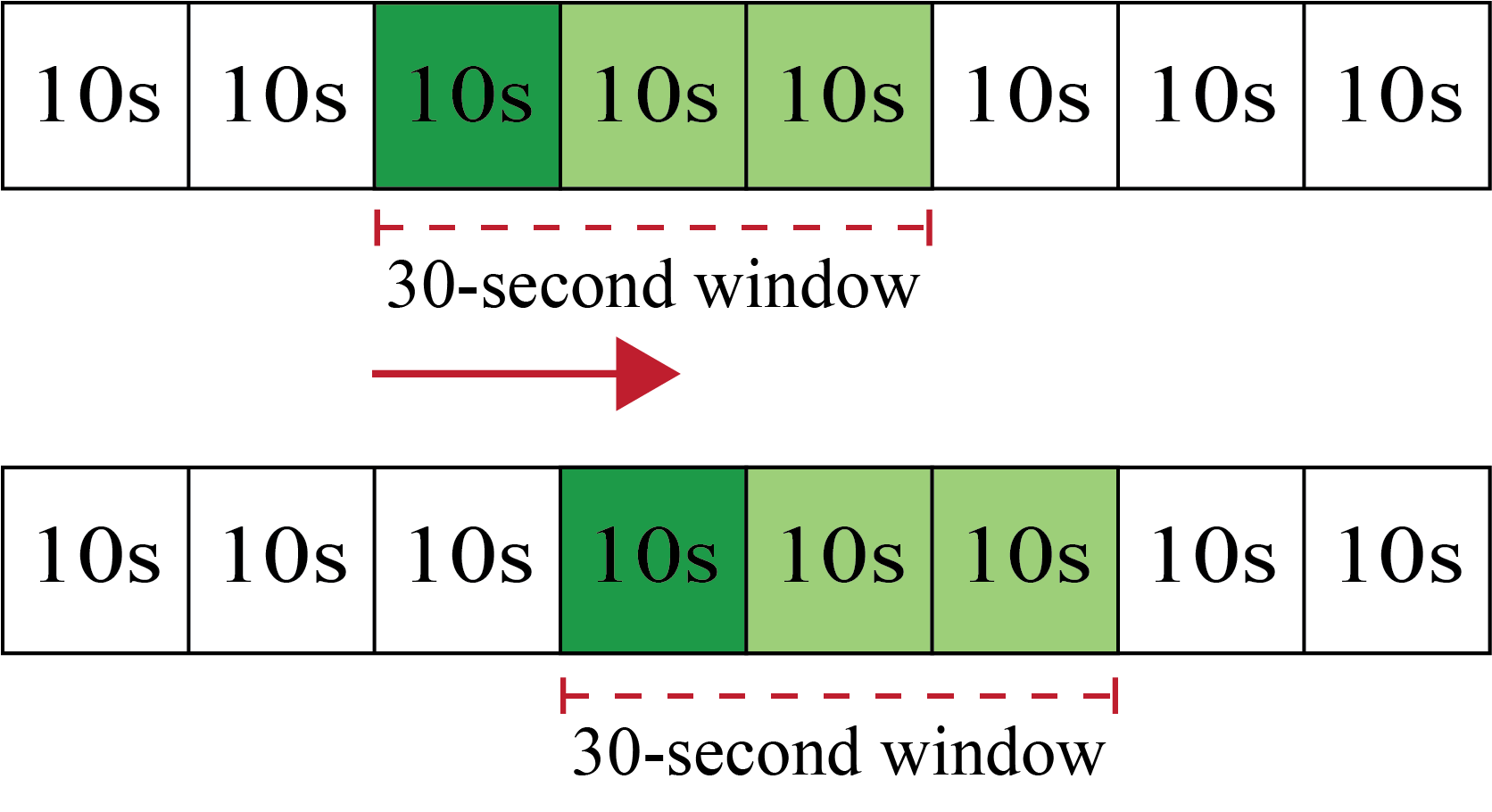}%
			\label{fig_second_case}}
		\caption{Windowing approaches: (a) 5-minute windows, 30-second steps for machine learning, (b) 30-second windows with 10-second steps for deep learning.}
		\label{fig_sim}
	\end{figure*}

		In this study, we introduce two windowing strategies for segmenting ECG signals, specifically designed for machine learning and deep learning models. These techniques are tailored to optimize the models for accuracy, temporal resolution, and computational efficiency, which are essential for practical wearable applications. The first windowing technique is for machine learning models, and the second for deep learning models. The windowing strategies used are illustrated in Figure~\ref{fig_sim}.
	
	\begin{enumerate}
		\item{\textbf{Machine Learning Windowing (5-minute window with 30-second step):} For the machine learning models, we employed a 5-minute window with a 30-second step, resulting in a 30-second temporal resolution. This extended window size is particularly important for capturing both frequency-domain and nonlinear-domain features, which are crucial for accurate classification and require at least 60 seconds of ECG signal to ensure reliability \cite{11}. Prior work often used 30-second segments in the middle of a 5-minute window for feature extraction \cite{2}, \cite{9}, \cite{28}, \cite{29}. However, we have placed the 30-second window at the start of the 5-minute segment rather than the middle, which led to an improvement in accuracy over previous methods. This strategic placement enhances the ability to detect early signal dynamics, which are key to distinguishing between sleep stages. This novel shift in window placement contributes to better performance, as seen in Figure~\ref{fig_first_case}, where the segmentation is depicted.}
		\item{\textbf{Deep Learning Windowing (30-second window with 10-second step):} 
		For deep learning models, we applied a 30-second window with a 10-second step, achieving a much finer temporal resolution of 10 seconds. This approach enables near real-time detection of sleep transitions, making it particularly suitable for wearable devices that need dynamic, responsive monitoring. Unlike traditional methods, this shorter window size improves the detection of transient events, such as micro-arousals or sudden sleep transitions, which are crucial for accurate sleep monitoring but are often missed with longer windowing strategies. The key advantage of this windowing strategy is its ability to balance temporal resolution with computational efficiency, reducing the processing power required without sacrificing accuracy. Due to automatic feature extraction in deep learning models, this 10-second window allows the model to achieve better accuracy than traditional methods that rely on longer windows. This technique is illustrated in Figure~\ref{fig_second_case}, where the segmentation is shown with the shorter window and smaller step size.}
	\end{enumerate}
	
The 30-second window with a 10-second step enables sleep stage prediction at a finer temporal resolution of 10 seconds while preserving sufficient physiological context from the ECG signal. This higher temporal granularity allows the system to respond more effectively to sleep stage transitions and transient events, which is particularly valuable for wearable sleep monitoring systems requiring responsive and energy-efficient operation.

It should be noted that overlapping windowing itself is a well-established technique in signal processing and is not claimed as the primary novelty of this work. Rather, the contribution lies in leveraging this windowing configuration to enable ECG-based 4-stage sleep stage classification at a 10-second temporal resolution while maintaining computational efficiency suitable for wearable deployment.

	\subsection{Feature Extraction and Selection}
	To capture sleep stage effects on the pattern of ECG signals, we extract a comprehensive set of features from both R waves or EDR and RR intervals or HRV. These features are grouped into four categories: EDR Features, HRV time-domain features, HRV frequency-domain features, and HRV non-linear features \cite{11}. We compute these features for both the 5-minute window and the 30-second step size.\\
	EDR features capture respiratory patterns from ECG signals, providing insights into breathing dynamics during sleep. We have extracted 18 features from EDR shown in Table~\ref{tab:feature_description} with brief descriptions. HRV time-domain features represent statistical and geometric measures of HRV, reflecting the overall cardiac autonomic balance. Table~\ref{tab:hrv_features} presents these features \cite{12}. HRV frequency-domain features in Table~\ref{tab:frequency_domain_features} are derived from spectral analysis of HRV, providing information about the distribution of signal power in different frequency bands, which correspond to various physiological processes \cite{12}. HRV non-linear domain features in Table~\ref{tab:complexity_variability_measures} capture complex, non-linear patterns in heart rate dynamics, offering insights into the intricate regulatory mechanisms of the cardiovascular system during sleep \cite{12}. In total, we have extracted these features from both 30-second steps and 5-minute windows. This dual approach results in 187 features for each timeframe, providing a comprehensive representation of ECG signal characteristics at different temporal resolutions. These features, derived from both the shorter 30-second steps and the longer 5-minute windows, serve as inputs for our machine learning models to classify sleep stages.
	
	To address the curse of dimensionality and optimize model performance, we employed the RFE algorithm for feature selection. This method iteratively removes less important features while evaluating model performance. Starting with all  187 extracted features, RFE gradually prunes the set, retaining only the 30 most influential attributes. This process helps prevent overfitting, improves model generalization, and reduces computational complexity by focusing on the most relevant ECG signal characteristics.

	The specific features selected by the RFE algorithm are reported here to improve transparency and reproducibility. Two features derived from the short-term windowing configuration with a 30-second step were selected, namely \texttt{HRV\_MeanNN} and \texttt{HRV\_Prc20NN}. The remaining selected attributes were extracted from the longer 5-minute window and include the following HRV-based and statistical features: \texttt{HRV\_MeanNN}, \texttt{HRV\_SDANN1}, \texttt{HRV\_CVSD}, \texttt{HRV\_MCVNN}, \texttt{HRV\_SDRMSSD}, \texttt{HRV\_Prc80NN}, \texttt{HRV\_pNN20}, \texttt{HRV\_MinNN}, \texttt{HRV\_MaxNN}, \texttt{HRV\_CSI\_Modified}, \texttt{HRV\_IALS}, \texttt{HRV\_PSS}, \texttt{HRV\_PI}, \texttt{HRV\_DFA\_alpha1}, \texttt{HRV\_ShanEn}, \texttt{HRV\_FuzzyEn}, \texttt{HRV\_CD}, \texttt{HRV\_HFD}, together with statistical descriptors including \texttt{max}, \texttt{mean}, \texttt{var}, \texttt{rmse}, \texttt{skewness}, \texttt{waveform\_factor}, \texttt{peak\_factor}, \texttt{impulse\_factor}, \texttt{margin\_factor}, and \texttt{rms}. This combination of HRV indices and statistical descriptors enables the model to capture both autonomic nervous system dynamics and statistical characteristics of the ECG-derived signals, supporting the physiological relevance of the selected feature set.
	
	\begin{table}[H]
		\centering
		\caption{EDR Features\label{tab:feature_description}}
		\renewcommand{\arraystretch}{1.5}
		\resizebox{\textwidth}{!}{%
			
			\begin{tabular}{
					>{\centering\arraybackslash}m{0.16\linewidth}
					>{\centering\arraybackslash}m{0.72\linewidth}
				}
				\toprule
				\textbf{Feature} & \textbf{Description} \\
				\midrule
				max, min, mean, median, std, var & Statistical measures capturing the distribution and variability of the EDR signal amplitudes \\
				\midrule
				peak\_to\_peak & Difference between maximum and minimum values \\
				\midrule
				rmse & Root Mean Square Error, a measure of signal fluctuations \\
				\midrule
				kurtosis & "Peakedness" of the amplitude distribution \\
				\midrule
				skewness & Asymmetry of the amplitude distribution \\
				\midrule
				waveform\_factor & Ratio of RMSE to mean, reflects waveform shape \\
				\midrule
				peak\_factor & Ratio of peak-to-peak to RMSE, indicates "spikiness" \\
				\midrule
				impulse\_factor & Ratio of peak-to-peak to mean, also reflects "spikiness" \\
				\midrule
				margin\_factor & Ratio of peak-to-peak to RMS, provides dynamic range information \\
				\midrule
				rms & Root Mean Square of the signal amplitudes \\
				\bottomrule
			\end{tabular}%
		}
	\end{table}
	
	\begin{table}[H]
		\centering
		\caption{The HRV Time Domain Features\label{tab:hrv_features}}
		\renewcommand{\arraystretch}{1.5}
		\resizebox{\textwidth}{!}{%
			
			\begin{tabular}{
					>{\centering\arraybackslash}m{0.16\linewidth}
					>{\centering\arraybackslash}m{0.72\linewidth}
				}
				\toprule
				\textbf{Feature} & \textbf{Description} \\
				\midrule
				MeanNN, SDNN, RMSSD, SDSD & Measures of overall heart rate variability and short-term variability. \\
				\midrule
				CVNN, CVSD & Coefficient of variation of RR intervals and standard deviation of successive differences. \\
				\midrule
				MedianNN, MadNN, MCVNN, IQRNN & Measures of central tendency and dispersion of RR intervals. \\
				\midrule
				SDRMSSD & Square root of the mean squared differences between successive RR intervals. \\
				\midrule
				Prc20NN, Prc80NN & Percentage of RR intervals that are less than 20\% and greater than 80\% of the mean RR interval. \\
				\midrule
				pNN50, pNN20 & Percentage of RR intervals that differ by more than 50 and 20 milliseconds, respectively. \\
				\midrule
				MinNN, MaxNN & Minimum and maximum RR intervals. \\
				\midrule
				HTI & Heart rate turbulence index. Measures the complexity of heart rate variability. \\
				\midrule
				TINN & Total index of nonlinearity. Measures the nonlinearity of heart rate dynamics. \\
				\bottomrule
			\end{tabular}%
		}
	\end{table}
	
	\begin{table}[H]
		\centering
		\caption{The HRV Frequency Domain Features\label{tab:frequency_domain_features}}
		\renewcommand{\arraystretch}{1.5}
		\resizebox{\textwidth}{!}{%
			
			\begin{tabular}{
					>{\centering\arraybackslash}m{0.16\linewidth}
					>{\centering\arraybackslash}m{0.72\linewidth}
				}
				\toprule
				\textbf{Feature} & \textbf{Description} \\
				\midrule
				LF (Low-Frequency) & Typically associated with sympathetic activity and stress responses. \\
				\midrule
				HF (High-Frequency) & Often linked to parasympathetic activity and respiratory rate. \\
				\midrule
				VHF (Very High-Frequency) & Related to microcirculation and baroreflex sensitivity. \\
				\midrule
				LFHF & Ratio of LF to HF. Indicates the balance between sympathetic and parasympathetic activity. \\
				\midrule
				LFn, HFn, LnHF & Normalized versions of LF, HF, and LFHF, respectively, to account for variations in heart rate. \\
				\bottomrule
			\end{tabular}%
		}
	\end{table}
	
	\begin{table}[H]
		\centering
		\caption{The HRV Non-linear Domain Features\label{tab:complexity_variability_measures}}
		\renewcommand{\arraystretch}{1.5}
		\resizebox{\textwidth}{!}{%
			
			\begin{tabular}{
					>{\centering\arraybackslash}m{0.20\linewidth}
					>{\centering\arraybackslash}m{0.80\linewidth}
				}
				\toprule
				\textbf{Feature} & \textbf{Description} \\
				\midrule
				SD1, SD2, SD1SD2 & Measures of short-term and long-term variability in heart rate. \\
				\midrule
				S, CSI, CVI & Measures of the complexity and irregularity of heart rate dynamics. \\
				\midrule
				CSI\_Modified, PIP, IALS, PSS, PAS & Modified complexity measures and measures of Poincare plot analysis. \\
				\midrule
				GI, SI, AI, PI & Geometric measures of the Poincare plot. \\
				\midrule
				C1d, C1a, SD1d, SD1a, C2d, C2a, SD2d, SD2a, Cd, Ca, SDNNd, SDNNa & Measures of variability in different time domains. \\
				\midrule
				DFA\_alpha1, MFDFA\_alpha1\_Width, MFDFA\_alpha1\_Peak, MFDFA\_alpha1\_Mean, MFDFA\_alpha1\_Max, MFDFA\_alpha1\_Delta, MFDFA\_alpha1\_Asymmetry, MFDFA\_alpha1\_Fluctuation, MFDFA\_alpha1\_Increment & Measures of long-range correlations and fractal properties of heart rate dynamics. \\
				\midrule
				ApEn, SampEn, ShanEn, FuzzyEn, MSEn, CMSEn, RCMSEn & Entropy measures that quantify the complexity and randomness of heart rate dynamics. \\
				\midrule
				CD, HFD, KFD, LZC & Measures of complexity, fractal dimension, and long-range correlations. \\
				\bottomrule
			\end{tabular}%
		}
	\end{table}

	\subsection{Machine Learning Algorithms}
	In this study, we employed six diverse machine learning algorithms to classify sleep stages based on features extracted from HRV and EDR signals, which were selected using the Recursive Feature Elimination (RFE) algorithm. While our primary focus is on energy-efficient deep learning, classical machine learning methods are inherently lightweight and can operate without a graphics processing unit (GPU), making them well-suited for microprocessors and wearable applications. However, these methods depend on feature extraction and typically require longer window sizes, which may limit their responsiveness in real-time settings. By incorporating these algorithms, we perform a comprehensive analysis to evaluate the trade-offs between classification performance and energy consumption. The algorithms used include K-Nearest Neighbor (KNN), Support Vector Machine (SVM), Logistic Regression (LR), Decision Tree (DT), Random Forest (RF), and Gradient Boosting Decision Tree (GBDT). Each offers a distinct approach to the classification task, ranging from instance-based learning to ensemble methods. The following subsections provide a brief overview of each algorithm and its application to our sleep stage classification problem. All models were trained using 5-fold cross-validation on the training set and evaluated on the test set to assess classification performance.
	
	\textit{\textbf{KNN}} is a non-parametric method that classifies data points based on how close it is to other samples in the feature space. For a new data point, KNN identifies the K closest samples from the training set and assigns the most common class among these neighbors as the prediction label \cite{30}.
	
	\textit{\textbf{SVM}} is a powerful algorithm that aims to find the optimal hyperplane to separate different classes in a high-dimensional space. SVM maximizes the margin between classes, which is the distance between the hyperplane and the nearest data points from each class \cite{31}.
	
	\textit{\textbf{LR}} is a straightforward and commonly used method for predicting the probability that an input belongs to a specific category. It takes input features, applies a mathematical formula to create a weighted sum, and then uses a sigmoid function to squeeze the result into a range between 0 and 1. This final value represents the probability of the input belonging to a particular class \cite{32}.
	
	\textit{\textbf{DT}} creates a tree-like model in which each internal node represents a decision based on a feature, and each leaf node represents a class label. The decision process involves moving from the root to the leaf nodes, following branches based on answers to questions, until a class label is reached \cite{33}.
	
	\textbf{\textit{RF}} is an ensemble learning method that constructs multiple independent parallel decision trees and combines their outputs to make predictions. Each tree is built using a random subset of features and training samples, which helps to reduce overfitting and improve generalization. The final classification is determined by a majority vote across all trees. This approach often leads to improved accuracy and robustness compared to single decision trees \cite{34}.
	
	\textbf{\textit{GBDT}} is an ensemble technique that constructs decision trees one after another (serial), each tree aiming to fix the errors made by the previous ones. Unlike RF, where trees are trained simultaneously, GBDT builds its trees in sequence and merges their outputs. This method uses gradient descent to reduce prediction errors \cite{35}.
	
	\subsection{Hyperparameter Tuning}
	Hyperparameters are crucial configuration settings in machine learning algorithms that must be set before training and can significantly influence model performance. Examples include the number of neighbors in KNN, the regularization strength in SVM, and the tree depth in Random Forests. To efficiently optimize these hyperparameters, we employed a Bayesian optimization framework, which uses internal 5-fold cross-validation as part of its probabilistic search strategy. Specifically, Bayesian optimization models the objective function based on past evaluations and uses this information to guide subsequent sampling of the hyperparameter space. This approach not only avoids exhaustive grid searches or random sampling but also provides a more reliable estimate of model performance through the built-in cross-validation, leading to faster convergence on near-optimal configurations.
	
	\subsection{Deep Learning Architectures}
	In this study, we explore eight deep learning architectures for sleep stage classification, each evaluated on 30-second ECG windows with a 10-second step, thereby achieving a 10-second resolution in detecting sleep transitions. Since deep learning models can be computationally demanding, we specifically focus on lightweight variants suitable for wearable applications, where power constraints and near real-time responsiveness are critical. Alongside these established architectures, we develop a SleepLiteCNN with a notably smaller parameter count, tailored to the unique characteristics of ECG-based sleep staging. The subsections that follow provide a concise overview of each architecture, highlighting their relative computational complexities and potential for low-power deployment.
	
	\textbf{\textit{VGG-11}}, the most compact version of the VGG family, was chosen for its well-established performance and straightforward, uniform structure \cite{36}. Our implementation of VGG-11 includes eight 1D convolutional layers, each with 3x1 filters, followed by ReLU activations and 2x1 max-pooling layers, concluding with three fully connected layers. This results in a model with 271 million parameters, making it a powerful yet computationally intensive option for feature extraction and classification.
	
	\textbf{\textit{MobileNet-v1}} is designed for mobile and resource-constrained platforms, using depthwise separable convolutions to minimize parameters and computational load \cite{37}. With only 400K parameters, it is highly suitable for wearable applications where energy efficiency is critical, making it ideal for real-time processing on low-power devices.
	
	\textbf{\textit{AlexNet}}, famous for its success in the 2012 ImageNet competition, consists of five convolutional layers followed by three fully connected layers, all using ReLU activations. It introduced important innovations like dropout and local response normalization, with our version containing 141 million parameters \cite{38}, making it an effective model for tasks like sleep stage classification from ECG data.
	
	\textbf{\textit{SqueezeNet-1.1}} specifically aims for AlexNet level accuracy with fewer parameters, making it ideal for deployment on devices with limited memory or power \cite{39}. It uses a "fire module" that combines 1x1 and 3x1 convolutions, reducing the model to just 354K parameters. This makes SqueezeNet highly efficient for applications that require both performance and low power consumption, such as wearable health monitoring devices.
	
	\textbf{\textit{Inception-v3}} uses "Inception modules" to run multiple convolutions with different filter sizes in parallel, allowing it to capture a wide range of features \cite{40}. Our version has 3.9 million parameters, offering strong performance in tasks where recognizing detailed patterns in data, like ECG signals, is important.
	
	\textbf{\textit{EfficientNet-B2}} is designed for efficiency, using mobile inverted bottleneck convolution blocks (MBConv) and squeeze-and-excitation techniques \cite{41}. With only 159K parameters, it delivers a good balance between performance and energy consumption, making it ideal for lightweight classification tasks in resource-limited environments.
	
	\textbf{\textit{ResNet-18}} is the smallest model in the ResNet family, with 17 1D convolutional layers and one fully connected layer. It uses residual connections to prevent the common training issue of vanishing gradients \cite{42}. With 3.8 million parameters, this architecture is robust yet efficient for applications needing lower computational power.
	
\begin{figure}[H]
		\centering
		\includegraphics[width=0.6\linewidth]{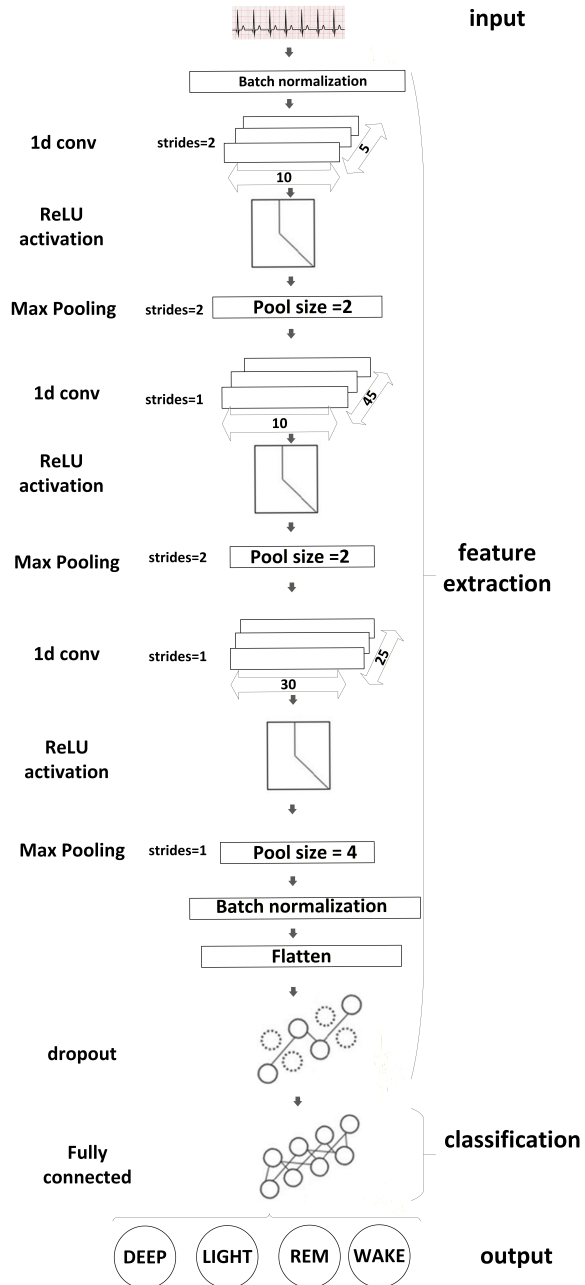}
		\caption{SleepLiteCNN Architecture.}
		\label{fig_3}
	\end{figure}

	\textbf{\textit{SleepLiteCNN}} While lightweight CNNs such as MobileNet and SqueezeNet offer a reduced computational load compared to larger architectures, they still include substantial parameter counts and lack an explicit optimization for ECG-based sleep stage classification. To address these limitations, we developed a SleepLiteCNN through a systematic analysis of ECG signal properties and the unique requirements of sleep staging (Figure~\ref{fig_3}).
	We have based our architecture design on three key principles: (1) effective feature extraction through progressive filter configurations, (2) minimal parameter counts for energy efficiency, and (3) robust generalization capability. We have evolved the architecture through iterative refinement and extensive experimentation, which has led us to design an efficient architecture comprising three convolutional layers with filter configurations of 5, 45, and 25 respectively. We have determined this progressive filter configuration through our empirical analysis of feature complexity in sleep-related ECG patterns. In the first layer, we have implemented 5 filters to capture fundamental waveform characteristics such as R-peak morphology and basic rhythm patterns. We have expanded to 45 filters in the second layer to extract detailed features of complex temporal relationships and sleep-specific ECG variations. We have then reduced to 25 filters in the final layer to distill the most relevant features while preventing overfitting.
	We have utilized max-pooling operations and ReLU activations in each convolutional layer, while we have implemented batch normalization at the input layer to address inter-subject variations and dropout (rate = 0.5) to enhance generalization.
	The decreasing number of filters in the final convolutional layer acts as a lightweight feature compression stage, reducing parameter count and mitigating overfitting while maintaining sufficient representational capacity for sleep stage classification. Overall, the resulting network comprises approximately 47K parameters, striking a balance between computational efficiency and classification accuracy. This lean design underscores the architecture's suitability for resource-constrained environments, where low power consumption is essential for continuous sleep monitoring. Notably, we have also applied the same SleepLiteCNN architecture in our prior work on sleep apnea subtype classification using ECG signals, further demonstrating its robustness and versatility for ECG-based biomedical classification tasks \cite{51}.
	
	\subsection{Energy Consumption Estimation, Quantization, and FPGA Synthesis}
	In this study, we applied quantization techniques to optimize energy consumption in deep learning models. Quantization reduces the precision of network weights and activations from 32-bit floating-point to lower bit-width representations \cite{13}, thereby decreasing both model size and computational requirements while largely preserving accuracy. We specifically chose 8-bit quantization, as this format is widely supported across most hardware platforms---including microcontroller units (MCUs), central processing units (CPUs), GPUs, and FPGAs---unlike other bit-widths (e.g., 6-bit or 10-bit), which are less commonly supported. To implement quantization, we used QKeras, a quantization-aware extension of Keras that also includes tools for estimating energy consumption at the 45 nm technology node \cite{13}. Quantization was applied to the convolutional, activation, and fully connected layers of our deep learning models.
	
	Based on balancing accuracy and energy efficiency, we selected our SleepLiteCNN model for FPGA synthesis---not only due to its favorable energy-accuracy trade-off, but also to validate its feasibility under hardware resource constraints. For this purpose, we used hls4ml, a tool that automates the conversion of machine learning models into FPGA implementations using High-Level Synthesis (HLS) \cite{14}. The model was synthesized on the xc7a200tfbg484-3 device from the Artix-7 FPGA family.
	
	\section{Results}
	\label{Results}
	This section presents our experimental results, which validate the primary contributions of our study. First, we demonstrate high-accuracy 4-stage sleep classification using ECG signals---with MobileNet-v1 achieving up to 92\% accuracy and 91\% F1-Score. We then highlight the effectiveness of our windowing strategies: a 5-minute window with a 30-second step for machine learning models to capture high-quality frequency and nonlinear-domain features, and a 30-second window with a 10-second step for deep learning models to enable near-real-time, 10-second resolution detection. Our comprehensive evaluation compares multiple machine learning algorithms and deep learning architectures on the same 4-stage classification task (wake, REM, light sleep, and deep sleep). Furthermore, by applying 8-bit quantization through quantization aware training, we significantly reduce inference energy consumption while maintaining accuracy. To address power constraints in wearable devices, we developed a SleepLiteCNN designed for energy-efficient inference and validated its feasibility by analyzing its resource utilization on an FPGA. Together, these results underscore the potential of our approach for achieving accurate, energy-efficient, and real-time sleep stage classification in wearable applications. 
	
	Given the class imbalance in our dataset, we use macro-averaged F1-score as the primary evaluation metric to ensure balanced performance across both minority and majority classes. We also report macro-averaged precision and recall to further validate model effectiveness across all sleep stages.
	
	Table~\ref{tab:model_performance} summarizes the performance of various machine learning models using a 5-minute window with a 30-second step to extract features from HRV and EDR signals for 4-stage sleep classification (wake, REM, light, deep). These features capture essential time, frequency and nonlinear characteristics crucial for distinguishing between sleep stages. Among the evaluated models, GBDT achieves the highest performance, with 89\% accuracy and a 89\% F1-score. We suspect that GBDT's iterative error correction enables it to effectively capture the complex nonlinearities inherent in ECG-derived features. RF and KNN also deliver robust performance, both registering accuracy and F1-scores in the 90-88\% range, underscoring the strength of ensemble and instance-based approaches for this task. All models were tuned using Bayesian optimization with internal 5-fold cross-validation. In contrast, SVM shows slightly lower performance with 86\% accuracy and F1-score, possibly due to its relatively limited capacity to model intricate data patterns. LR, relying on linear decision boundaries, performs poorly with 60\% accuracy and a 49\% F1-score, reflecting its inability to capture the complex variations present in the features. Additionally, a standalone DT achieves only 78\% accuracy and a 77\% F1-score, further highlighting the challenges faced by simpler models. Overall, these findings suggest that methods capable of handling nonlinear relationships---such as GBDT, RF, and KNN---are better suited for ECG-based sleep staging using our feature extraction approach.

	\begin{table}[H]
		\centering
		\caption{Performance Comparison of Machine Learning Models\label{tab:model_performance}}
		\renewcommand{\arraystretch}{1.5}
		\resizebox{\textwidth}{!}{%
			
			\begin{tabular}{
					>{\centering\arraybackslash}m{0.14\linewidth}
					>{\centering\arraybackslash}m{0.2\linewidth}
					>{\centering\arraybackslash}m{0.2\linewidth}
					>{\centering\arraybackslash}m{0.2\linewidth}
					>{\centering\arraybackslash}m{0.2\linewidth}
				}
				\toprule
				\textbf{Model} & \textbf{Accuracy (\%)} & \textbf{F1-Score (\%)} & \textbf{Precision (\%)} & \textbf{Recall (\%)} \\
				\midrule
				GBDT & 89 & 89 & 91 & 87 \\
				\midrule
				RF & 90 & 89 & 91 & 88 \\
				\midrule
				KNN & 88 & 88 & 89 & 88 \\
				\midrule
				SVM & 86 & 86 & 86 & 86 \\
				\midrule
				DT & 78 & 77 & 76 & 77 \\
				\midrule
				LR & 60 & 49 & 59 & 46 \\
				\bottomrule
			\end{tabular}%
		}
	\end{table}

	Table~\ref{tab:cnn_performance} presents the performance of various deep learning models, all evaluated using our  30-second windowing approach with a 10-second step, which achieves near-real-time, 10-second resolution---a capability not demonstrated by previous methods. This strategy highlights the advantage of automatic feature extraction over traditional methods. Among the models, MobileNet-v1 achieves up to 92\% accuracy and 91\% F1-score  with an approximate 5\% improvement in accuracy and a 9\% improvement in F1-score over previous ECG-based methods \cite{2}. Its lightweight architecture, built on depthwise separable convolutions, appears particularly well-suited for extracting salient features from short ECG segments. VGG-11 and AlexNet also demonstrate strong performance, each achieving over 90\% accuracy and F1-score, suggesting that deeper networks can effectively capture complex patterns in ECG data when appropriately configured. Our SleepLiteCNN, designed specifically for energy-efficient inference, records competitive performance with 90\% accuracy and an 89\% F1-score, striking a balance between classification performance and resource efficiency. By contrast, SqueezeNet-1.1 (84\% accuracy, 84\% F1-score), Inception-v3, EfficientNet-B2, and ResNet-18 (42\%---77\% accuracy range) exhibit lower performance. One likely reason is that these architectures---originally tailored to 2D image tasks---may not optimally align with the distinctive morphology and temporal dependencies of 1D ECG signals, even when adapted. Inception and ResNet blocks, for example, can excel at hierarchical feature extraction in images but may prove less effective for the shorter, rapidly changing segments characteristic of ECG-based sleep staging. Furthermore, over-parameterization in certain deep networks (like Inception-v3) could lead to suboptimal generalization when trained on smaller ECG datasets. Overall, these findings highlight that deep learning models can effectively exploit shorter window sizes to achieve high classification performance, clearly surpassing traditional machine learning in temporal resolution. Notably, MobileNet-v1 and our SleepLiteCNN illustrate the potential for robust accuracy with streamlined architectures, thus paving the way for near-real-time, energy-efficient ECG-based sleep stage classification in wearable applications.
	
	To further quantify the reliability and statistical stability of the reported results, we estimated the performance variability of the two best-performing models-MobileNet-v1 and SleepLiteCNN-using a non-parametric bootstrap procedure on the held-out test set. Confidence Intervals (CI; specifically 95\% bootstrap confidence intervals) provide a statistical range within which the true performance metric is expected to lie, reflecting how sensitive the model's results are to sampling variability in the test data. To compute these intervals, we generated 1000 bootstrap resamples of the test samples (sampling with replacement), recalculated the overall accuracy and macro F1-score for each resample, and derived the mean and the 2.5th-97.5th percentiles of the resulting distributions. This procedure offers a principled, training-free method for estimating uncertainty and is particularly useful in physiological signal analysis, where test sets are often limited in size and model outputs may exhibit non-Gaussian variability. The resulting estimates show that MobileNet-v1 achieves an accuracy of 91.2\% (95\% CI: 90.6-91.6\%) and a macro F1-score of 91.4\% (95\% CI: 90.9-91.9\%), indicating both high accuracy and tight statistical bounds. Similarly, the proposed SleepLiteCNN achieves an accuracy of 88.9\% (95\% CI: 88.4-89.5\%) and a macro F1-score of 88.4\% (95\% CI: 87.8-88.9\%), demonstrating stable performance despite its lightweight architecture. These narrow confidence intervals collectively confirm that both models provide robust and statistically consistent classification performance on the held-out test set.	
	\begin{table}[H]
		\centering
		\caption{Performance Comparison of Deep Learning Models\label{tab:cnn_performance}}
		\renewcommand{\arraystretch}{1.5}
		\resizebox{\textwidth}{!}{%
			\footnotesize
			\begin{tabular}{
					>{\centering\arraybackslash}m{0.14\linewidth}
					>{\centering\arraybackslash}m{0.15\linewidth}
					>{\centering\arraybackslash}m{0.15\linewidth}
					>{\centering\arraybackslash}m{0.15\linewidth}
					>{\centering\arraybackslash}m{0.15\linewidth}
				}
				\toprule
				\textbf{Model} & \textbf{Accuracy (\%)} & \textbf{F1-Score (\%)} & \textbf{Precision (\%)} & \textbf{Recall (\%)} \\
				\midrule
				SleepLiteCNN & 89 & 89 & 89 & 88 \\
				\midrule
				MobileNet-v1 & 92 & 91 & 91 & 91 \\
				\midrule
				VGG-11 & 91 & 90 & 93 & 87 \\
				\midrule
				AlexNet & 91 & 91 & 91 & 91 \\
				\midrule
				SqueezeNet-1.1 & 84 & 84 & 83 & 84 \\
				\midrule
				Inception-v3 & 77 & 74 & 76 & 73 \\
				\midrule
				EfficientNet-B2 & 75 & 73 & 73 & 74 \\
				\midrule
				ResNet-18 & 42 & 40 & 50 & 45 \\
				\bottomrule
			\end{tabular}%
		}
	\end{table}
	
	We applied 8-bit quantization to the convolutional, activation, and fully connected layers of all deep learning models. Table~\ref{tab:model_comparison} compares the estimated energy consumption and performance of these models before and after quantization. Our SleepLiteCNN demonstrated the lowest energy consumption, requiring just 79.74~$\mu J$ in full precision and only 5.48~$\mu J$ after quantization, while improving from 89\% accuracy and 89\% F1-score to 90\% in both metrics post-quantization.
	
	MobileNet-v1, although lightweight in design, consumed significantly more energy than SleepLiteCNN---494.94~$\mu J$ in full precision and 30.84~$\mu J$ after quantization. However, it delivered the highest classification performance in full precision, achieving 92\% accuracy and 91\% F1-score, with a modest reduction to 89\% and 88\%, respectively, after quantization. This makes MobileNet-v1 a suitable choice when performance is prioritized over ultra-low energy usage.
	
	Larger models such as VGG-11 and AlexNet exhibited high energy demands (up to 8.4~mJ), yet showed only minor performance gains. Interestingly, SqueezeNet-1.1 and Inception-v3 both saw improved accuracy after quantization---despite lower parameter counts---highlighting a key insight: quantization-aware training does not necessarily degrade performance and can, in some cases, enhance it. This improvement likely results from weight re-scaling and regularization effects introduced during quantization-aware training, which help adapt model parameters more effectively to 8-bit representations.
	
	EfficientNet-B2, while relatively energy-efficient, showed limited performance and a slight drop in F1-score after quantization. ResNet-18, with its high parameter count, remained the least effective in both accuracy and energy metrics, despite showing an unexpected performance gain after quantization.
	
	In Table~\ref{tab:model_comparison}, an increase in accuracy is observed for ResNet-18 after applying 8-bit quantization. Although quantization is commonly associated with a slight reduction in performance, such improvements can occur when quantization-aware training  introduces additional regularization during optimization. By constraining weights and activations to lower-precision representations, quantization-aware training effectively injects quantization noise into the training process, which can reduce overfitting and improve generalization. This effect can be particularly noticeable for relatively overparameterized models such as ResNet-18 when trained on datasets of limited size. Consequently, the quantized model may occasionally achieve higher accuracy than its full-precision counterpart.
	
	Overall, 8-bit quantization consistently reduced energy consumption across all evaluated models, and in several cases preserved or even enhanced classification performance. These results validate the effectiveness of quantization-aware training for developing energy-efficient deep learning models tailored to resource-constrained environments like wearable health monitoring.
	
	\begin{table}[H]
		\caption{Energy Consumption and Performance Comparison of Deep Learning Models Before and After 8-bit Quantization.}
		\label{tab:model_comparison}
		\centering
		\renewcommand{\arraystretch}{1.5}
		
		\resizebox{\textwidth}{!}{%
			\begin{tabular}{cccccccc}
				\toprule
				&  & \multicolumn{3}{c}{\textbf{Full Precision}}              & \multicolumn{3}{c}{\textbf{8-bit Quantization}}          \\ 
				\cmidrule{3-5} \cmidrule{6-8}
				\textbf{Model} & \textbf{Total Params} & \textbf{Accuracy (\%)} & \textbf{F1-Score (\%)} & \textbf{Energy ($\mu$J)} & \textbf{Accuracy (\%)} & \textbf{F1-Score (\%)} & \textbf{Energy ($\mu$J)} \\ \midrule
				SleepLiteCNN     & 47K                  & 89                    & 89                     & 79.74                & 90                    & 90                     & 5.48                 \\ \midrule
				MobileNet-v1   & 400K                 & 92                    & 91                     & 494.94               & 89                    & 88                     & 30.84                \\ \midrule
				VGG-11         & 271M                 & 91                    & 90                     & 8402.94              & 87                    & 86                     & 7346.02              \\ \midrule
				AlexNet        & 141M                 & 91                    & 91                     & 2631.26              & 89                    & 89                     & 2158.74              \\ \midrule
				SqueezeNet-1.1 & 354K                 & 84                    & 84                     & 477.84               & 89                    & 89                     & 477.34               \\ \midrule
				Inception-v3   & 3.9M                 & 77                    & 74                     & 2193.47              & 78                    & 74                     & 2187.88              \\ \midrule
				EfficientNet-B2 & 159K                & 75                    & 73                     & 12.21                & 75                    & 71                     & 11.98                \\ \midrule
				ResNet-18      & 3.8M                 & 42                    & 40                     & 3019.03              & 58                    & 48                     & 191.11               \\ \bottomrule
		\end{tabular}}
		
	\end{table}
	
	Based on the balance between accuracy and energy efficiency, we selected our SleepLiteCNN model for FPGA synthesis---not only because it demonstrated the lowest energy consumption and a strong energy-accuracy trade-off, but also to showcase the feasibility of our self-designed architecture in terms of hardware resource utilization. We synthesized the model on the xc7a200tfbg484-3 device from the Artix-7 FPGA family. Table~\ref{tab:Utilization} summarizes the FPGA resource utilization for our SleepLiteCNN before and after applying 8-bit quantization. Notably, quantization led to significant reductions in resource usage for example, Look-Up Table (LUT) usage decreased from 38\% to 30\%, and flip-flop usage dropped from 56\% to 47\% confirming that our SleepLiteCNN is not only energy-efficient but also highly suitable for deployment on resource-constrained platforms.

	\begin{table}[H]
		\caption{FPGA Resource Utilization for SleepLiteCNN Model Before and After 8-bit Quantization\label{tab:Utilization}}
		\centering
		\renewcommand{\arraystretch}{1.5}
		
		\begin{tabular}{ccc}
			\toprule
			& \multicolumn{2}{c}{\textbf{Utilization (\%)}}    \\ 
			\cmidrule{2-3}
			\textbf{Resource}  & \textbf{Full Precision}  &  \textbf{8-bit Quantization} \\ \midrule
			LUT & 38 & 30 \\ \midrule
			FF & 56 & 47 \\ \midrule
			BRAM & 28 & 24 \\ \midrule
			DSP & 35 & 22 \\ \bottomrule
		\end{tabular}
	\end{table}

\begin{figure}[H]
		\centering
		\subfloat[]{\includegraphics[width=0.7\linewidth]{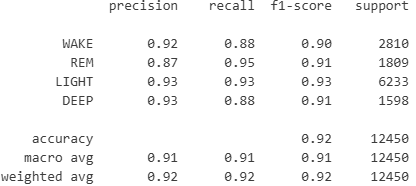}%
			\label{mob1}}
		\vfil
		\subfloat[]{\includegraphics[width=0.6\linewidth]{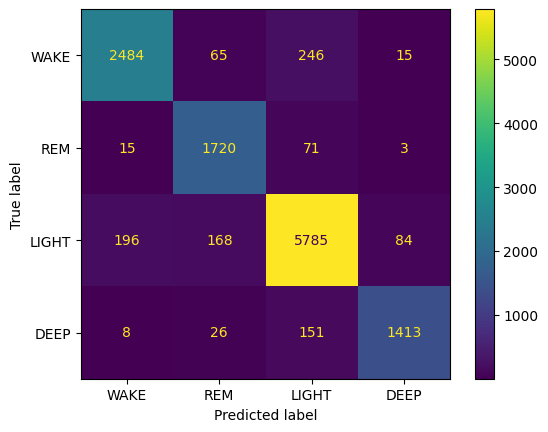}%
			\label{mob2}}
		
		\caption{Classification report (a) and confusion matrix (b) for MobileNet-v1, our best-performing model for sleep stage classification.}
		\label{mobile}
	\end{figure}

	To provide a more detailed per-class analysis, Figure~\ref{mobile} presents the classification report and confusion matrix for MobileNet-v1, our best-performing model for sleep stage classification. The results demonstrate MobileNet-v1's high accuracy and consistent performance across all four sleep stages, with strong precision, recall, and F1-scores---particularly for the challenging REM and deep sleep categories.
	
\begin{figure}[H]
		\centering
		\subfloat[]{\includegraphics[width=0.7\linewidth]{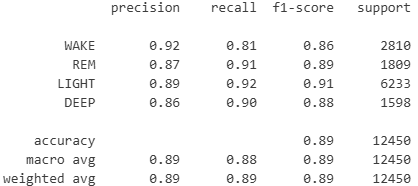}%
			\label{cnn1}}
		\vfil
		\subfloat[]{\includegraphics[width=0.6\linewidth]{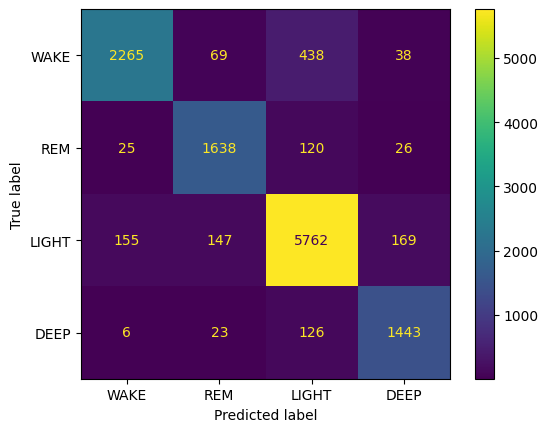}%
			\label{cnn2}}
		\vfil
		\centering
		\subfloat[]{\includegraphics[width=1.1\linewidth]{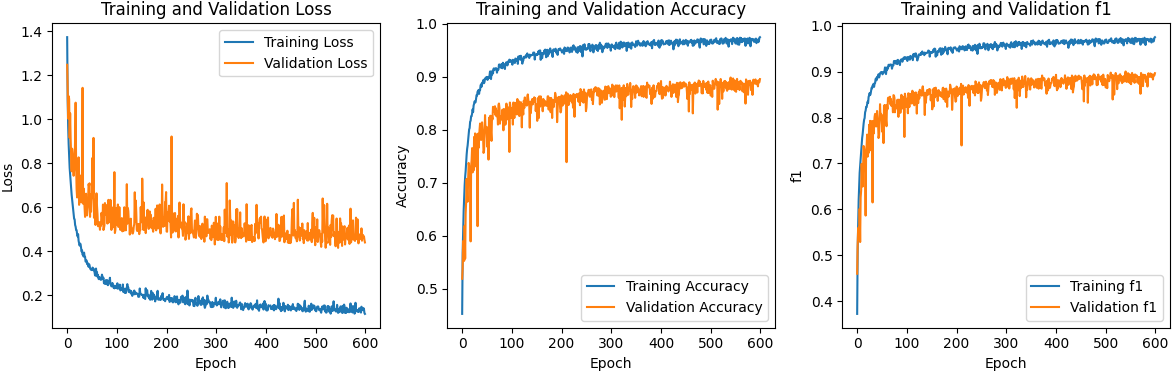}%
			\label{cnn_v}}
		
		\caption{Performance evaluation of the SleepLiteCNN model: (a) classification report and (b) confusion matrix before quantization; (c) Training and validation curves for loss, accuracy, and F1-score over 600 epochs.}
		\label{cnn}
	\end{figure}

	In Figure ~\ref{cnn}, we present the performance of the SleepLiteCNN model in full precision (32-bit). Subfigure (a) shows the classification report, where the model achieves 89\% overall accuracy and 89\% macro-averaged F1-score. The model performs reliably across all four sleep stages, with particularly strong results in the LIGHT and DEEP stages. Subfigure (b) displays the confusion matrix, which reveals some misclassification between WAKE and LIGHT stages---a common challenge in sleep staging due to overlapping ECG characteristics. Subfigures (c) show the training dynamics over 600 epochs. The model was trained for 600 epochs to ensure stable convergence; as shown in the training curves, the loss, accuracy, and F1-score exhibit smooth convergence with closely aligned training and validation trends. The loss, accuracy, and F1-score curves demonstrate smooth convergence and consistent validation trends, confirming effective learning and generalization.
	
\begin{figure}[H]
		
		\centering
		\subfloat[]{\includegraphics[width=0.7\linewidth]{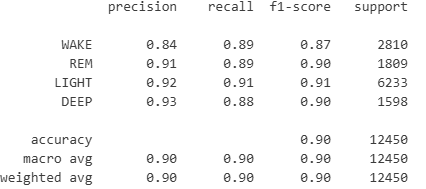}%
			\label{cnnq1}}
		\vfil
		\subfloat[]{\includegraphics[width=0.6\linewidth]{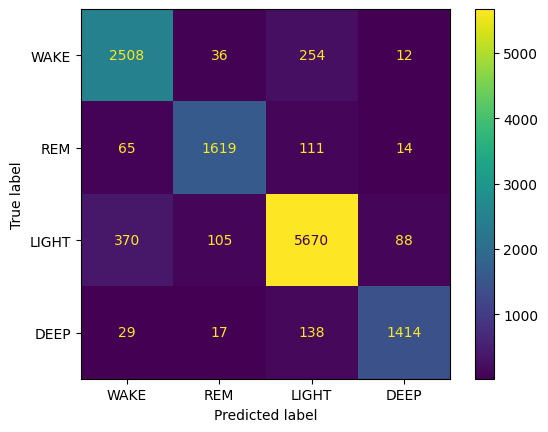}%
			\label{cnnq2}}
		
		\vfil
		\centering
		\subfloat[]{\includegraphics[width=1.1\linewidth]{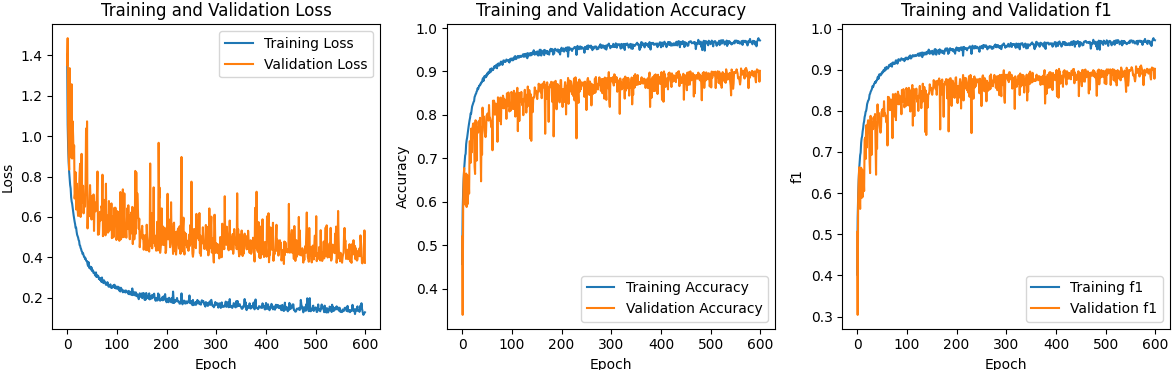}%
			\label{cnn_vq}}
		\caption{Performance evaluation of the SleepLiteCNN model after 8-bit quantization: (a) classification report and (b) confusion matrix before quantization; (c)Training and validation curves for loss, accuracy, and F1-score over 600 epochs.}
		\label{cnn_q}
	\end{figure}

	Figure ~\ref{cnn_q} presents the performance of the SleepLiteCNN model after 8-bit quantization. The classification report in subfigure (a) indicates slightly improved accuracy and F1-score (both 90\%), particularly in the REM and DEEP stages. The confusion matrix in subfigure (b) shows more balanced predictions compared to the full-precision version. Subfigures (c) illustrate that the training and validation trajectories remain stable after quantization, with no signs of degraded learning dynamics.
	
	These findings confirm that quantization-aware training preserves---and in some cases improves---model performance, by enabling more efficient use of 8-bit weight representations. The SleepLiteCNN model thus remains highly effective and robust, making it a strong candidate for low-power, real-time sleep stage classification in wearable devices.
	
	We compare our proposed ECG-based sleep staging approach with similar state-of-the-art methods in the literature, focusing on (1) multi-stage classification performance and (2) energy consumption. Table~\ref{tab:sleep_stage_models} summarizes representative works on 4-stage sleep classification, emphasizing the temporal resolution, signal modality, and performance metrics (accuracy and F1-score). Table~\ref{tab:sleep_stage_models_en} highlights the energy consumption of our approach relative to state-of-the-art hardware implementations. In particular, we detail how we scaled our 45~nm results to 180~nm to enable a fair comparison with older CMOS technology nodes used by prior studies.
	
	A survey of Table~\ref{tab:sleep_stage_models} reveals that our best-performing deep learning model, MobileNet-v1, achieves an accuracy of 92\% and an F1-score of 91\% for 4-stage classification (wake, REM, light, deep), with an approximate 5\% improvement in accuracy and a 9\% improvement in F1-score over previous ECG-based methods \cite{2}. Crucially, this high performance is obtained at a 10-second effective resolution, implemented using a 30-second window with a 10-second step size, thereby enabling near--real-time detection of sleep transitions. In contrast, most prior methods employ considerably longer windows---from 30 seconds to 5 minutes---raising the risk of missing clinically important, short-duration events such as micro-arousals. Moreover, our approach relies solely on ECG signals, making it more practical for wearable applications than strategies that require multiple sensors (such as EEG or respiratory inductive plethysmography). The significant improvement in both accuracy and F1-score likely stems from our use of automatic feature extraction within a deep learning framework, coupled with a novel short-window approach that effectively preserves the temporal specificity of the ECG data.
	
	A further comparison is provided in Table \ref{tab:sleep_stage_models_en}, 
	contrasting the estimated energy consumption of our two primary deep learning
	models (MobileNet-v1 and our SleepLiteCNN) with that of existing systems 
	implemented on integrated circuits at the 180 nm node. Because our 
	measurements are based on a 45 nm process, we scale these energy figures 
	to approximate their equivalents at 180 nm by applying a conservative
	scaling factor of about 50. This factor accounts for differences in transistor 
	geometry, supply voltage, and parasitic capacitances between the 45 nm and 180 nm nodes.

	The results indicate that our MobileNet-v1 model, after 8-bit quantization, consumes roughly 30.84\,$\mu$J at 45 nm---corresponding to about 1.6 mJ at 180 nm. In contrast, our SleepLiteCNN, which is specifically optimized for ECG signals and resource efficiency, requires only 5.48\,$\mu$J per inference at 45 nm, or roughly 0.28 mJ when scaled to 180 nm. Even after scaling, these energy consumption figures compare very favorably with previous works, which report energy usage ranging from 0.15 mJ to over 10 mJ for similar or even coarser temporal resolutions in EEG-based designs.

	This comparison highlights several key points. First, our approach clearly improves energy efficiency compared to older CMOS implementations---even after accounting for process scaling to 180 nm. Second, achieving a near-real-time, 10-second resolution does not incur a significant energy penalty, unlike some earlier designs that rely on EEG signals and more complex hardware. Finally, using ECG instead of EEG or EMG contributes to a more patient-friendly solution, as ECG requires fewer electrodes and can be captured using simpler wearable devices.
	
	Combining the results from Tables \ref{tab:sleep_stage_models} and \ref{tab:sleep_stage_models_en}, we conclude that our 8-bit quantized MobileNet-v1 and SleepLiteCNN not only exceed prior methods in accuracy and temporal resolution but also maintain substantially lower power requirements when scaled to older technologies. This balance of performance and energy efficiency makes our method a promising candidate for continuous sleep staging in compact, battery-powered devices.
	
			It should be noted that the energy scaling from 45 nm to 180 nm in Table \ref{tab:sleep_stage_models_en} intended as an approximate comparison to provide a general estimate of relative energy consumption. In practical hardware implementations, the actual energy usage may vary depending on factors such as supply voltage, memory access patterns, and operating clock frequency.
	
	\begin{table}[H]
		\centering
		\caption{Comparison of Our Approach with State-of-the-Art Sleep Stage Classification Models}
		\label{tab:sleep_stage_models}
		\renewcommand{\arraystretch}{1.5}
		
		\resizebox{\textwidth}{!}{%
			\footnotesize 
			\begin{tabular}{>{\centering\arraybackslash}m{0.12\linewidth}>{\centering\arraybackslash}m{0.08\linewidth}>{\centering\arraybackslash}m{0.14\linewidth}>{\centering\arraybackslash}m{0.19\linewidth}>{\centering\arraybackslash}m{0.09\linewidth}>{\centering\arraybackslash}m{0.1\linewidth}>{\centering\arraybackslash}m{0.09\linewidth}}
				\toprule
				\textbf{Model} & \textbf{Signal} & \textbf{Temporal Resolution} & \textbf{Class Labels} & \textbf{Dataset} & \textbf{Accuracy (\%)} & \textbf{F1-Score (\%)} \\
				\midrule
				LSTM \cite{9} & ECG & 30 Second (5-minute window) & WAKE, REM, Stage 2, SWS & 1,441 patients & 77 & - \\ \midrule
				CNN \cite{10} & ECG & 30-second & WAKE, REM, LIGHT, DEEP & SHHS, MESA, CinC & 78, 80, 72 for each dataset & - \\ \midrule
				SVM \cite{6} & ECG & 5-minute segments & WAKE, REM, Stage 2, SWS & SHHS & 89 & - \\ \midrule
				BLDC \cite{5} & ECG, RIP & 30-Second (various window sizes) & WAKE, REM, LIGHT, DEEP & SIESTA & 69 & - \\ \midrule
				SVM \cite{4} & ECG, RIP, MOV & 60-second & WAKE, REM, N1+N2, N3 & 85 PSG & 69 & - \\ \midrule
				GBDT \cite{2} & ECG & 30 Second (5-minute window) & WAKE, REM, LIGHT, DEEP & UCDDB, MIT-BIH & 87 & 82 \\ \midrule
				CNN \cite{44} & 2 EEG , 1 EOG & 30 Second (150-Second window) & WAKE, N1, N2, N3, REM & Sleep-EDF & 81 & 72 \\ \midrule
				DCNN \cite{45} & 2 EEG & 30 Second (180-Second window) & WAKE, S1+S2, S3+S4, REM & Sleep-EDFX & 91.27 & 90.88 \\ \midrule
				Bi-LSTM \cite{46} & EEG & 30 Second & WAKE, N1, N2, N3, REM & Sleep-EDF & 81.6 & 74.7 \\ \midrule
				SleepGCN \cite{47} & EEG, EOG & 30 Second & WAKE, N1, N2, N3, REM & SleepEDF-20 & 89.70 & 85.20 \\ \midrule
				SleepContextNet \cite{49} & EEG & 30-Second (240-Second window) & WAKE, N1, N2, N3, REM & Sleep-EDF & 84.8 & 79.8 \\ \midrule
				
				MixSleepNet \cite{50} & EEG, EOG, EMG, ECG & 30 Second & WAKE, N1, N2, N3, REM & ISRUC-S3 & 83.0 & 82.1 \\ \midrule

				\textbf{Our best model (MobileNet-v1)} & ECG & \textbf{10-Second (30-second window)} & WAKE, REM, LIGHT, DEEP & UCDDB & \textbf{92} & \textbf{91} \\ \midrule
				\textbf{Our SleepLiteCNN (after 8-bit Quantization)} & ECG & \textbf{10-Second (30-second window)} & WAKE, REM, LIGHT, DEEP & UCDDB & \textbf{90} & \textbf{90} \\ \bottomrule
		\end{tabular}}
	\end{table}
	\begin{table}[H]
		\centering
		\caption{Comparison of Estimated Energy Consumption: Our ECG-Based Approach vs. Existing Methods}
		\label{tab:sleep_stage_models_en}
		\renewcommand{\arraystretch}{1.5}
		
		\resizebox{\textwidth}{!}{%
			\footnotesize
			\begin{tabular}{>{\centering\arraybackslash}m{0.16\linewidth}>{\centering\arraybackslash}m{0.05\linewidth}>{\centering\arraybackslash}m{0.15\linewidth}>{\centering\arraybackslash}m{0.12\linewidth}>{\centering\arraybackslash}m{0.095\linewidth}>{\centering\arraybackslash}m{0.13\linewidth}>{\centering\arraybackslash}m{0.14\linewidth}}
				\toprule
				\textbf{Model} & \textbf{Signal} & \textbf{Temporal Resolution} & \textbf{Class Labels} & \textbf{Accuracy (\%)} & \textbf{CMOS Technology} & \textbf{Energy} \\
				\midrule
				decision tree \cite{44} & 2 EEG & 30-second & W, N1, N2, N3, REM & 78.9 & 180 nm & 10.35 mJ \\
				\midrule
				NN-based decision tree \cite{17} & EEG + EMG & 30-second & W, N1 \& N2, N3, REM & 81 & 180 nm & 0.149 mJ \\
				
				\midrule
				Our best model (MobileNet-v1) after 8-bit Quantization & ECG & 10-Second (30-second window) & WAKE, REM, LIGHT, DEEP & 91 & 45 nm & 30.84 $\mu$J (1.6 mJ in 180 nm) \\
				\midrule
				Our SleepLiteCNN after 8-bit Quantization & ECG & 10-Second (30-second window) & WAKE, REM, LIGHT, DEEP & 90 & 45 nm & 5.48 $\mu$J (0.28 mJ in 180 nm) \\
				\bottomrule
		\end{tabular}}
	\end{table}

	\section{Limitations and Future Work}
	\label{Limitations}
	
	Despite the promising results achieved by the proposed SleepLiteCNN framework, several limitations should be acknowledged. First, the experiments were conducted exclusively on the UCDDB dataset, which contains recordings from only 25 subjects. Although this dataset provides high-quality ECG recordings and reliable sleep stage annotations, the relatively small cohort size limits the assessment of inter-subject generalization.
	
	In preliminary experiments, we explored a strict subject-independent validation strategy using Leave-One-Subject-Out (LOSO) cross-validation. However, due to the limited number of subjects in the dataset, a substantial performance degradation was observed, which highlights the challenges of training robust subject-independent models on small cohorts. This limitation suggests that larger and more diverse datasets are necessary to reliably evaluate generalization across unseen individuals.
	
	Future work will therefore focus on multi-dataset evaluation by combining UCDDB with other publicly available ECG-based sleep staging datasets. Such an approach would increase subject diversity and enable more reliable subject-independent validation protocols.
	
	In addition, the proposed SleepLiteCNN architecture is intentionally designed to be lightweight and computationally efficient, making it suitable for deployment on resource-constrained wearable devices. In practical wearable scenarios, models are often trained or fine-tuned using data from the target user, enabling personalized sleep monitoring while preserving data privacy. In such personalized settings, subject-specific physiological patterns can be effectively captured, and the compact architecture of SleepLiteCNN makes it particularly suitable for on-device inference.
	
	Future studies will therefore investigate both cross-dataset generalization and personalized on-device training strategies using ECG signals collected from real-world wearable devices.

	\section{Discussion}
	\label{Discussion}
	This study introduces a comprehensive framework for ECG-based 4-stage sleep classification that achieves a balance between high accuracy, near-real-time temporal resolution, and energy efficiency, addressing key limitations in prior wearable sleep monitoring systems. Our evaluation of lightweight deep learning models, such as MobileNet-v1, demonstrated superior performance with 92\% accuracy and 91\% F1-score, marking a 5\% improvement in accuracy and a 9\% in F1-score compared to the leading ECG-based method~\cite{2}. This advancement is largely attributable to our  windowing strategies: the 5-minute window with 30-second steps for machine learning, which enhances feature robustness in frequency and nonlinear domains, and the 30-second window with 10-second steps for deep learning, enabling unprecedented 10-second resolution without compromising predictive power. Furthermore, the custom-designed SleepLiteCNN, optimized for ECG signals with only 47K parameters, maintains strong performance (89\% accuracy and 89\% F1-score) while drastically reducing computational demands. The application of 8-bit quantization further refines this efficiency, yielding 5.48~$\mu$J per inference at 45~nm technology with no loss in accuracy (90\% post-quantization), as validated through FPGA synthesis. These results underscore the potential of our approach to enable continuous, unobtrusive sleep tracking in resource-constrained wearables, outperforming traditional methods that rely on longer epochs or multi-sensor setups.
	
	The 10-second temporal resolution enabled by the proposed framework may provide additional clinical value compared to conventional 30-second epoch-based sleep staging. Finer temporal granularity can improve the detection of rapid transitions between sleep stages, which are often associated with sleep instability and fragmentation. Such transitions may occur over short periods and can be partially obscured when predictions are limited to 30-second intervals. Higher temporal resolution may also facilitate the identification of brief arousals or micro-arousals that contribute to sleep fragmentation, which is commonly observed in disorders such as insomnia and sleep apnea. Furthermore, when combined with the energy-efficient design of the proposed system, this capability may support continuous long-term monitoring in wearable devices, enabling more detailed assessment of sleep dynamics in home-monitoring scenarios.

	In addition to energy efficiency, inference latency is also an important factor for near real-time wearable monitoring. Due to the lightweight architecture of SleepLiteCNN and the use of 8-bit quantization, the computational complexity of the model is relatively low. Consequently, the inference time on typical embedded or edge AI hardware is expected to be significantly shorter than the 10-second prediction update interval used in this work. Therefore, the model can provide sleep-stage updates in near real time without introducing meaningful processing delay.

	\section{Conclusion}
	\label{Conclusion}
	This study presents a practical and energy-efficient solution for near-real-time sleep stage classification using only ECG signals, addressing the limitations of traditional sleep monitoring methods. By introducing  windowing strategies, we successfully balance classification accuracy, temporal resolution, and energy consumption. While lightweight deep learning models like MobileNet-v1 achieve state-of-the-art performance, our custom-designed SleepLiteCNN optimized with 8-bit quantization---reduces energy consumption to just 5.48 $\mu J$ per inference, making it exceptionally well-suited for resource-constrained, wearable applications. Its successful deployment on an FPGA further confirms its feasibility for real-world, battery-operated devices.
	
	These findings lay the groundwork for more accessible and continuous sleep monitoring solutions. Future research will focus on validating the approach across larger and more diverse datasets and exploring adaptive quantization techniques to enhance hardware portability. By combining high-performance classification with resource-conscious implementation, this work contributes to the development of next-generation wearable health monitoring systems. Ultimately, this work bridges the gap between high-resolution sleep analytics and practical, long-term wearable deployment.
	
	\section*{Code Availability}
	The implementation code is publicly available at:  \href{https://github.com/zahraaayii/Energy-Efficient-Real-Time-4-Stage-Sleep-Classification-at-10-Second-Resolution}{github.com/zahraaayii/Energy-Efficient-Real-Time-4-Stage-Sleep-Classification-at-10-Second-Resolution}.

	\section*{Conflict of Interest}
	The authors declare that they have no known competing financial interests or personal relationships that could have appeared to influence the work reported in this paper.

	\bibliography{sn-bibliography}

	\section*{Biographies}
	
	\noindent
	\begin{minipage}{0.15\textwidth}
		\includegraphics[width=1in,height=1.25in,clip,keepaspectratio]{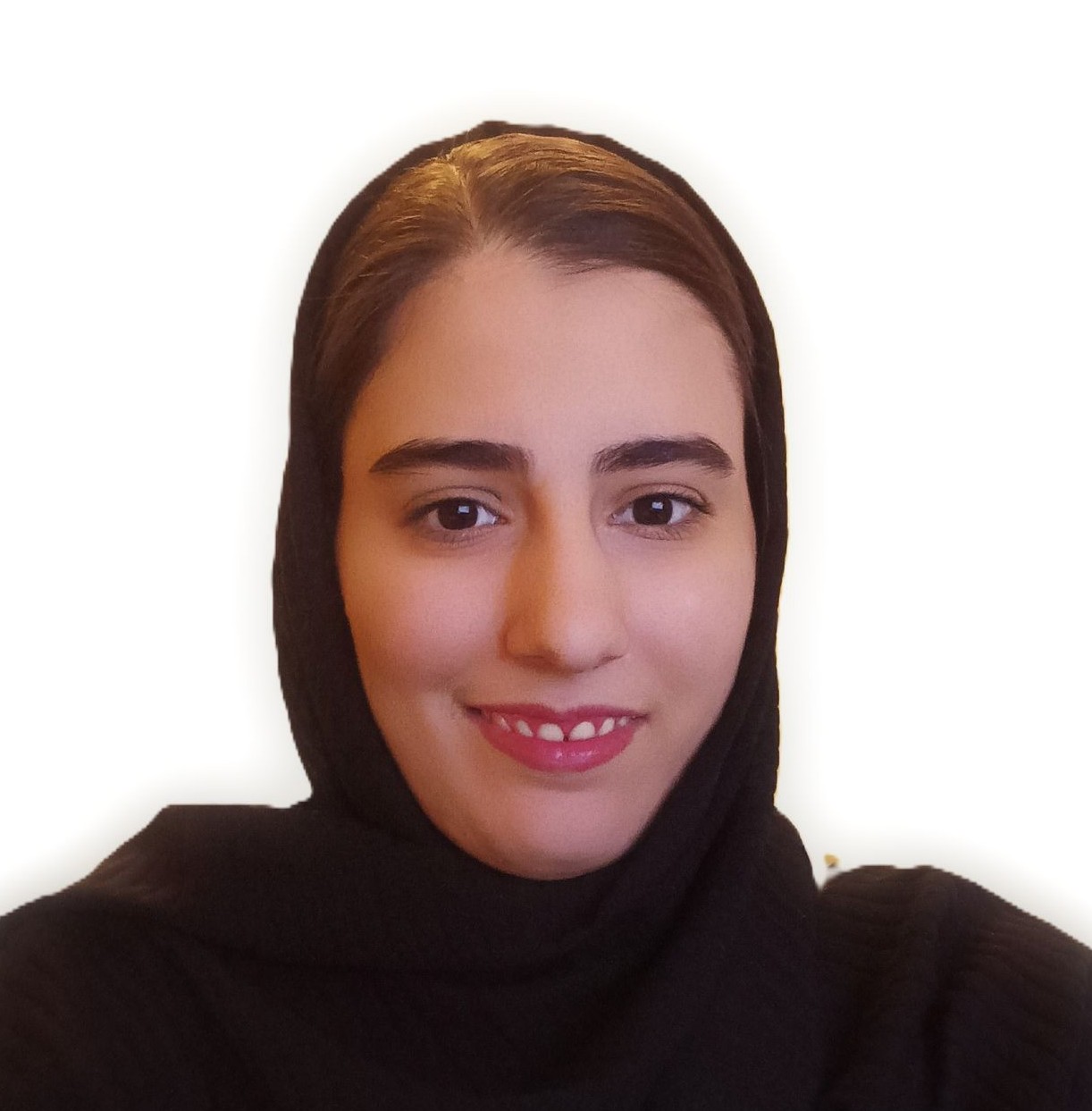}
	\end{minipage}%
	\hfill
	\begin{minipage}{0.8\textwidth}
		\textbf{Zahra Mohammadi} 	is currently a Ph.D. candidate in Computer Engineering at the University of Tehran, specializing in Computer Architecture. Her research primarily focuses on detecting sleep disorders through advanced analysis of biomedical signals, utilizing machine learning and deep learning techniques. She received her M.S. in Computer Engineering from the University of Tehran in 2023, where she ranked first among all graduates, and her B.S. in Computer Engineering from Shahid Beheshti University in 2020. 
	\end{minipage}
	
	\vspace{1em} 
	
	\noindent
	\begin{minipage}{0.15\textwidth}
		\includegraphics[width=1in,height=1.25in,clip,keepaspectratio]{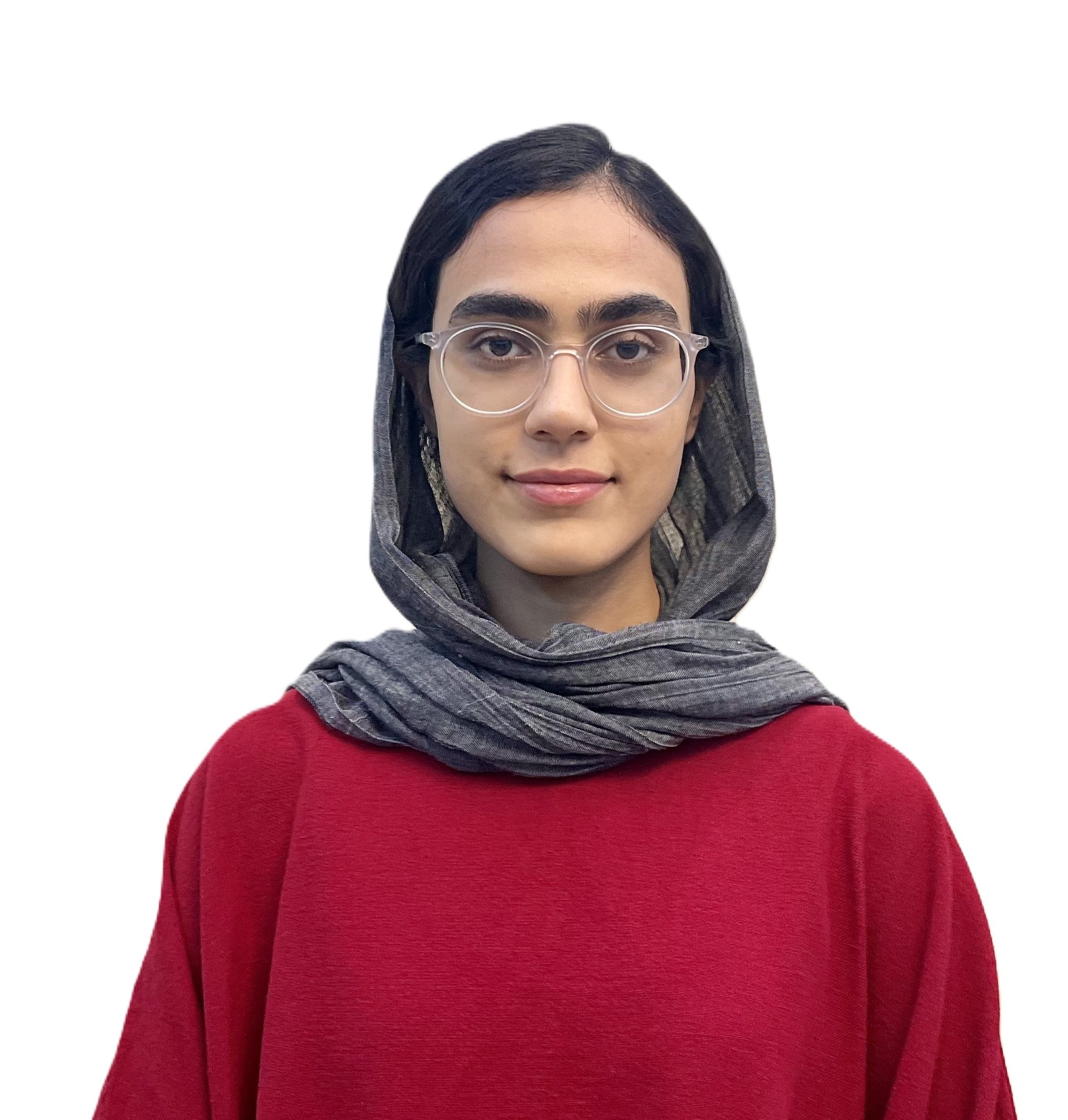}
	\end{minipage}%
	\hfill
	\begin{minipage}{0.8\textwidth}
		\textbf{Parnian Fazel} received the B.Sc. degree in Computer Engineering from the University of Tehran, Iran, in 2024. Her bachelor's thesis focused on automated sleep stage classification using machine learning techniques, for which she received the Best Undergraduate Project Award from the Department of Electrical and Computer Engineering at the University of Tehran. Her research interests include machine learning and deep learning, particularly their applications in the biomedical domain.
	\end{minipage}
	
	\vspace{1em}
	
	\noindent
	\begin{minipage}{0.15\textwidth}
		\includegraphics[width=1in,height=1.25in,clip,keepaspectratio]{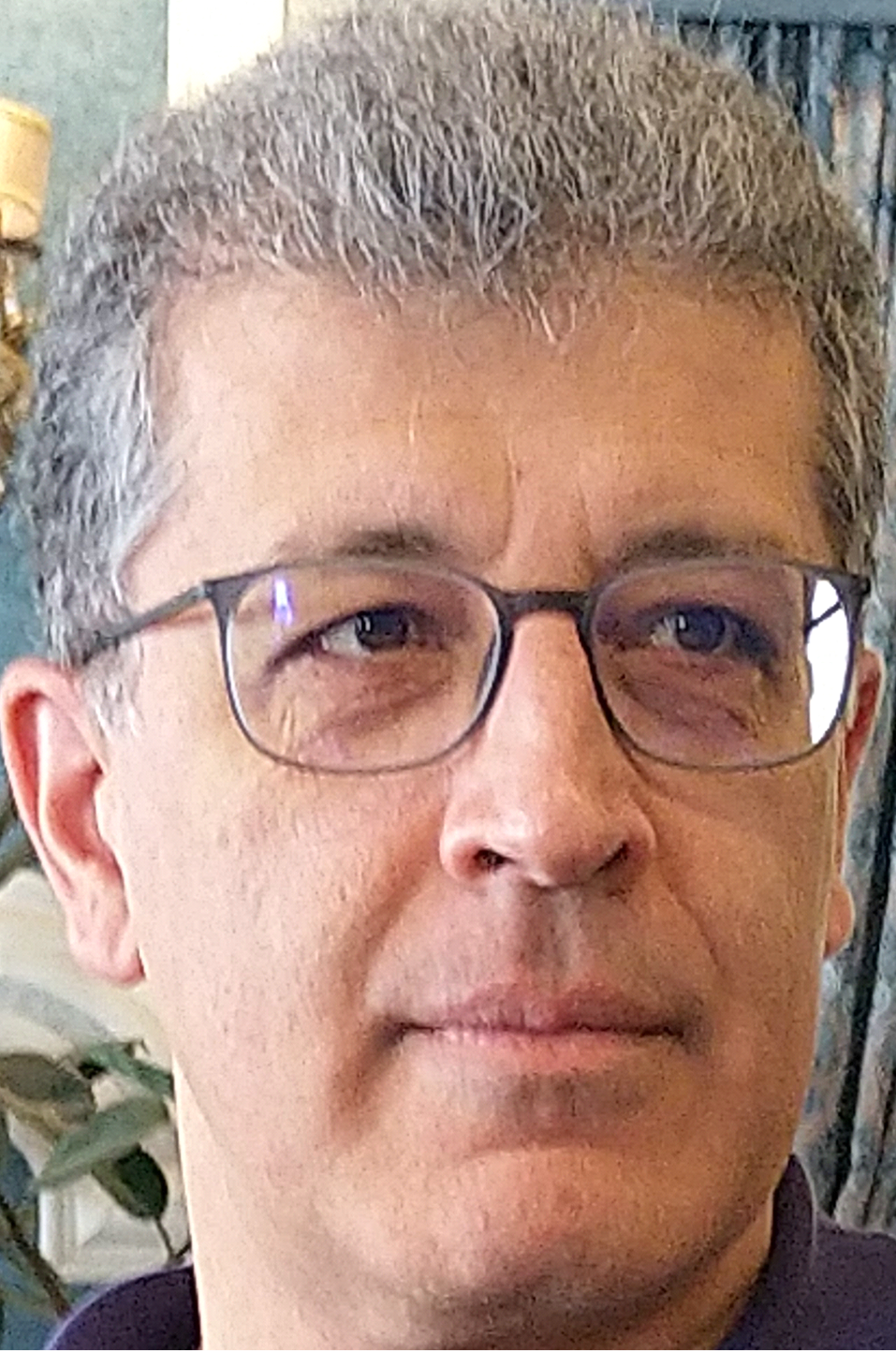}
	\end{minipage}%
	\hfill
	\begin{minipage}{0.8\textwidth}
		\textbf{Siamak Mohammadi} received his BSc, MSc and PhD degrees from the University of Paris Sud Orsay, France in 1990, 1992 and 1996, respectively, all in electrical engineering. From 1997 to 1999 he was a Research Associate with the Department of Computer Science, University of Manchester, England. In 1999 he moved to Canada and worked in Semiconductor industry in Toronto until 2005. Currently he is an Associate Professor in School of Electrical and Computer engineering, at the University of Tehran, Iran. He has over 30 years of experience in Low Power Design , Verification,  and Hardware Security of Embedded Systems.
	\end{minipage}

\end{document}